\documentclass[conference]{IEEEtran}

\usepackage{etoolbox}
\makeatletter
\patchcmd{\@makecaption}
  {\scshape}
  {}
  {}
  {}
\makeatletter
\patchcmd{\@makecaption}
  {\\}
  {.\ }
  {}
  {}
\makeatother

\usepackage{cite}
\usepackage{amsmath,amssymb,amsfonts}
\usepackage{graphicx}
\usepackage{multirow}
\usepackage{textcomp}
\usepackage{xcolor}
\usepackage{mathtools}
\usepackage{float}
\usepackage[utf8]{inputenc}
\usepackage{xparse}
\usepackage{hyperref}
\ifCLASSOPTIONcompsoc \usepackage[caption=false,font=normalsize,labelfont=sf,textfont=sf]{subfig}
\else
\usepackage[caption=false,font=footnotesize]{subfig}
\fi

\usepackage[ruled,lined, noend,linesnumbered]{algorithm2e}
\usepackage{algorithmicx}

\def\BibTeX{{\rm B\kern-.05em{\sc i\kern-.025em b}\kern-.08em
    T\kern-.1667em\lower.7ex\hbox{E}\kern-.125emX}}
    
\begin{document}

	\title{Flexible Communication for Optimal Distributed Learning over Unpredictable Networks}

	\author{\IEEEauthorblockN{Sahil Tyagi}
		\IEEEauthorblockA{\textit{Indiana University Bloomington, USA} \\
		styagi@iu.edu}
		\and
		\IEEEauthorblockN{Martin Swany}
		\IEEEauthorblockA{\textit{Indiana University Bloomington, USA} \\
		swany@indiana.edu}
	}
	
	\maketitle
	
	\begin{abstract}
		Gradient compression alleviates expensive communication in distributed deep learning by sending fewer values and its corresponding indices, typically via Allgather (AG).
Training with high compression ratio (CR) achieves high accuracy like DenseSGD, but has lower parallel scaling due to high communication cost (i.e., parallel efficiency).
Using lower CRs improves parallel efficiency by lowering synchronization cost, but degrades model accuracy as well (statistical efficiency).
Further, speedup attained with different models and CRs also varies with network latency, effective bandwidth and collective op used for aggregation.
In many cases, collectives like Allreduce (AR) have lower cost than AG to exchange the same amount of data.
In this paper, we propose an AR-compatible Top\textit{k} compressor that is bandwidth-optimal and thus performs better than AG in certain network configurations.
We develop a flexible communication strategy that switches between AG and AR based on which collective is optimal in the current settings, and model the pareto-relationship between parallel and statistical efficiency as a multi-objective optimization (MOO) problem to dynamically adjust CR and accelerate training while still converging to high accuracy.
	\end{abstract}
	
	\section{Introduction}\label{sec:intro}

Deep Neural Network (DNN) training requires numerous passes over the entire dataset (i.e., epochs) to achieve decent accuracy.
Synchronous Data-Parallel training \emph{scales-out} by processing different samples concurrently across multiple workers and computing local updates, which are then aggregated either via Parameter Server (PS) \cite{b0} or Allreduce (AR) \cite{b42}.
Gradient synchronization cost is further exacerbated by the growing size of DNNs, which causes slowdowns and degrade the \emph{parallel efficiency} of distributed training \cite{b1}.
For e.g., storing gradients as single-precision (32-bit) floats in a billion parameter model implies transmitting 4GB of updates on each iteration.
Communication collectives like Allgather (AG), Ring-based AR and Tree-based AR offer different latency-bandwidth trade-offs for the same message size \cite{b2}, which affects overall communication cost.
The bandwidth cost of sending-receiving data to-from PS increases linearly with cluster-size, while Ring-AR is bandwidth-optimal.
However, latency cost is low for PS and high in Ring-AR.
Further, latency or bandwidth itself can vary in a cluster over time due to various factors like resource sharing, QoS priorities, contention among multiple jobs, etc.
Thus, parallel efficiency of distributed training is also affected by network availability.

\vspace{0.1cm}
Gradient compression expedites training by lowering the exchange volume to reduce communication time, and is measured by the \emph{degree of compression} or \emph{Compression Ratio (CR)}.
High CR converge to high accuracy like DenseSGD (i.e., training \emph{without} any compression), albeit with little to no speedup.
Low CRs considerably reduce communication cost, but may degrade final accuracy or require more steps to converge, thus increasing training time \cite{b1}.
A majority of compression methods use AG to aggregate values and corresponding indices of compressed tensors across workers \cite{b2}.
In some network configurations, cost of AG can be high than using ring or tree AR to transmit the same message-size.
Thus, distributed training exhibits varying communication costs depending on DNN and cluster size under constrained networks with variable latency or bandwidth.
Even small models like ResNet18 \cite{b32} can suffer from frequent communication in high-latency, low-bandwidth settings.
Additionally, distributed DNN training suffers from \emph{statistical inefficiency} from noisy gradients generated by training with smaller mini-batches \cite{b3, b4}.
Noise is further aggravated when modela are updated by a compressed representation of the original gradients \cite{b1}.

\vspace{0.1cm}
We propose a novel compression technique called \textbf{AR-Top\textit{k}} \footnote{Code available at \href{https://github.com/sahiltyagi4/ARTopk}{https://github.com/sahiltyagi4/ARTopk}} that is compatible with both ring and tree-AR, and describe its two variants: \emph{staleness-based} (\textbf{STAR-Top\textit{k}}) and \emph{gradient variance-based} (\textbf{VAR-Top\textit{k}}), followed by their communication cost analysis.
We then develop heuristics to choose the optimal collective between AR-Top\textit{k} (ring or tree) and AG for a given network configuration.
To balance the pareto-relationship between parallel and statistical efficiency, we approach gradient compression as a multi-objective optimization problem to calculate optimal CR at different training stages.
	\section{Background And Related Work}\label{sec:bgrelated}

We first discuss the synchronization cost of data-parallel training as we scale from intra to inter-node devices, followed by studying how intra-node parallel efficiency is affected by latency, bandwidth, collective op used, model and cluster-size.
We also gloss over prior art covering statistical efficiency in distributed training and its impact on convergence quality/time.

\subsection{\textbf{Distributed Data-Parallel Training}}\label{sub:ddp}

Synchronous distributed training improves throughput by performing more work per-iteration.
With weak scaling, $N$ workers concurrently process mini-batch $\mathit{d_{(i, n)}}$ of size $b$ from dataset $\mathcal{D}_{n}$ to update model parameters $w_{i+1}$ at step $i$ by step-size $\eta$ to minimize loss function $\mathcal{L}(\cdot)$.
Eqn. (\ref{eqn:distsgd}) shows how the local updates are aggregated.
The compute cost of each iteration is identical, comprised of mini-batch sampling, computing and averaging gradients, followed by model update.
It is thus mainly comprised of computation, communication and IO overhead, shown in Eqn. (\ref{eqn:itrtime}).

\begin{subequations}
	\begin{equation}
		w_{i+1} = w_{i} - \eta \dfrac{1}{N} \sum_{n=1}^{n=N} \dfrac{\partial}{\partial w_{i}} (\dfrac{1}{|b|} \sum_{\mathit{d_{(i, n)}} \in \mathcal{D}_{n}} \mathcal{L}(x_{(i, n)},w_{i}))
		\label{eqn:distsgd}
	\end{equation}
	\begin{equation}
		t_{step} = t_{compute} + t_{sync} + t_{IO}
		\label{eqn:itrtime}
	\end{equation}
\end{subequations}

Fig. (\ref{computesynctime}) plots the computation and communication time of ResNet18, ResNet50 \cite{b32}, AlexNet \cite{b33} and Vision Transformer (ViT) \cite{b34} on single node with 8 NVIDIA V100 GPUs.
As DNN size increases (left to right), so does the communication overhead.
Compared to compute time, communication tends to be the primary bottleneck on modern ML accelerators.

\vspace{0.25cm}
\subsubsection{Intra vs. Inter-node Scaling}\label{subsub:interintranode}

 \begin{figure}[t]
	\subfloat[Compute vs. Sync Time]{\includegraphics[width=0.23\textwidth]{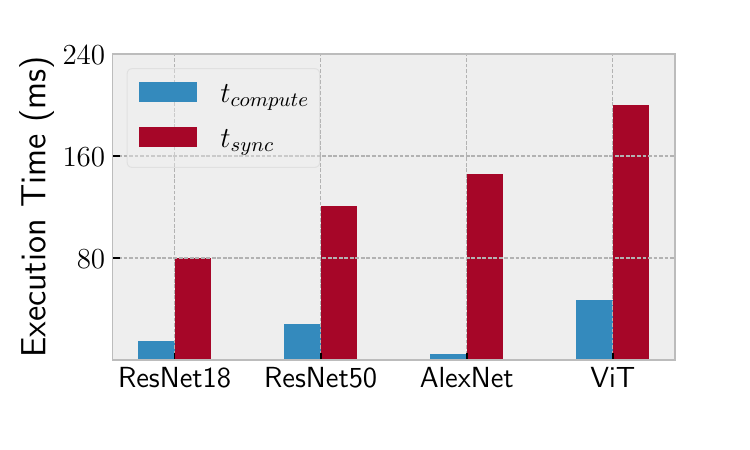}
	\label{computesynctime}}
	\hfill
	\subfloat[Inter vs. Intra-node Latency]{\includegraphics[width=0.23\textwidth]{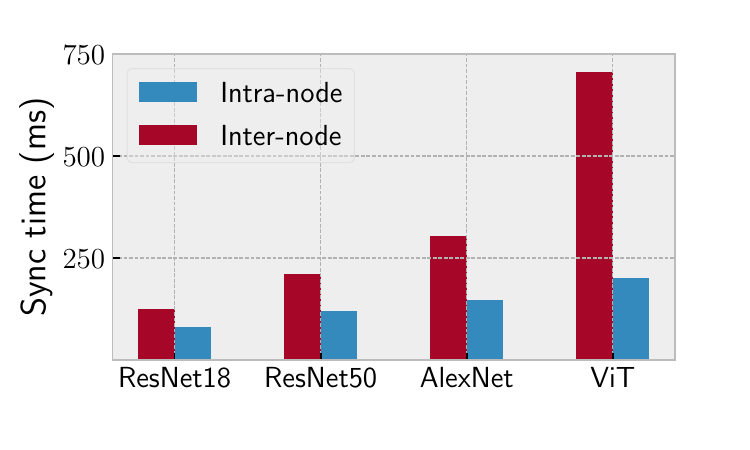}
	\label{interintranode}}
	\caption{(a) Intra-node computation-communication times for 8 workers. (b) Latency of gradient aggregation over 8 GPUs connected as: 8 GPUs/node vs. 1 GPU/node.
	Inter-node GPUs connected over 10Gbps network.}	
	\label{fig:syncbgfigs}
\end{figure}

In data-parallel training, synchronization cost to aggregate updates between nodes depends on the physical distance between them.
GPU devices on a single node are connected via high-speed interconnects like PCIe Gen4 (64GB/s), Gen5 (128GB/s) or NVLink 4.0 \cite{b6} with 900GB/s bandwidth to enable \emph{intra-node scaling}.
This type of scale-up on a single system is limited by finite PCIe lanes, for e.g., NVIDIA DGX systems can accommodate up to 8 GPUs/node \cite{b7} and NVSwitch can connect up to 256 GPUs in a cluster.
\emph{Inter-node scaling} spans out to multiple servers, but suffers from high communication cost to transmit massive tensors over networks with comparatively lower bandwidths.
The typical bandwidth in a data-center is about 10Gbps \cite{b9} and even lower at the edge.
We compare gradient synchronization costs for various DNNs in intra and inter-node settings.
In the former, 8 V100s are co-located on a single node (i.e., 8 GPUs/node) with peer-to-peer (P2P) transport enabled to allow CUDA direct access, while the latter is comprised of 8 machines with 1 GPU each connected via 10Gbps interface.
Inter-node communication cost is much higher while aggregating updates via Allreduce on NVIDIA's NCCL communication backend (Fig. (\ref{interintranode})).

\vspace{0.25cm}
\subsubsection{Cost of Communication Collectives}\label{subsub:alphabetacost}

\begin{table}
	\caption{Bandwidth Complexity and Communication Cost of Primitives}
	\centering
	\begin{tabular}{|c|c|c|c|c|}
		\hline
		\bfseries Operation & \bfseries BW Complexity & \bfseries Communication cost \\
		\hline
		PS (Star) & $\mathcal{O}(MN)$ & $2\alpha + 2(N-1)M\beta$ \\
		\hline
		Ring-AR & $\mathcal{O}(M)$ & $2(N-1)\alpha + 2\frac{(N-1)}{N}M\beta$ \\
		\hline
		Tree-AR & $\mathcal{O}(M\log (N))$ & $2\alpha\log (N) + 2\log (N) M\beta$ \\
		\hline
		Broadcast & $\mathcal{O}(M \log (N))$ & $\alpha \log (N) + \log (N) M\beta$ \\
		\hline
		Allgather & $\mathcal{O}(MN)$ & $\alpha \log (N) + (N-1)M\beta$ \\
		\hline
	\end{tabular}
\label{table:comcostops}
\end{table}

Gradients can be aggregated in various ways depending on the topology and algorithm used.
Centralized mechanisms exchange updates via Parameter servers (PS) based on a star-topology, while decentralized approaches use different flavors of Allreduce or AR (e.g., reduce-broadcast, scatter-reduce-allgather, ring, tree, etc.). 
Table (\ref{table:comcostops}) lists down the $\alpha$-$\beta$ communication cost \cite{b31, b42} and bandwidth complexity of collectives commonly used in distributed training.
The $\alpha$ term is latency, $1/\beta$ is bandwidth and $M$ is the model/gradient size exchanged among $N$ workers.
In PS, synchronization cost increases with both cluster and model-size.
Ring-AR is dominated by latency cost as its bandwidth-optimal, i.e., $\beta$ term is nearly independent of $N$ and increases only with $M$.
Although Tree-AR is \emph{logarithmic} bandwidth-optimal, it has lower latency cost than ring; $\alpha$ term is proportional to $\log (N)$ in the former vs. $(N-1)$ in the latter.
Hence, communication cost of different collectives varies for a given network configuration, cluster-size and model-size, as well as their implementation.
For e.g., AR over CUDA GPUs is faster with NCCL over Gloo, while Gloo performs better than MPI on CPUs.

\vspace{0.25cm}
\subsubsection{Communication Scheduling}\label{subsub:commsched}

Efficient communication scheduling accelerates training by hiding synchronization costs.
Although data-parallel training can be scaled out by processing data on many workers simultaneously, backpropagation itself is inherently sequential.
Input is fed to the innermost layer while outputs are generated on the outermost layer of DNNs.
By chain rule, gradients are first computed on the outer layers and propagate back to the inner layers.
ML frameworks improve iteration time by hiding communication cost and overlapping it with computation \cite{b11}, although this can be inefficient in DNNs with many layers of non-uniform size.
Thus, updates of current layer are sent while gradients of subsequent layer are still being computed.
With \emph{tensor-fusion} \cite{b18}, smaller layers are batched together to perform fewer communication operations.
Frameworks like PyTorch allow \emph{bucketing} to specify message-size for each allreduce call, with the default being 25MB.
Gradient computation time has decreased with the development of optimized GPU architectures, while size of DNNs continues to grow exponentially.
Thus, gain of interleaving computation with communication is saturating as the former tends to be lower than the latter.
\subsection{\textbf{Statistical Efficiency of Data-Parallel Training}}\label{sub:statefficiency}

Although deep learning is an iterative-convergent process, features learned by a model over the epochs are not uniformly attributed to each training step due to the stochastic nature of gradient descent, i.e., iterations do not contribute equally towards the overall learning of a model.
Gradients calculated from a randomly sampled mini-batch tend to be noisier than those computed over the full-batch (i.e., entire training data) \cite{b3}.
The divergence between local and averaged gradients represents the signal-to-noise ratio in the approximated gradients, which has prior been studied for fine-tuning cluster and batch-size in distributed training \cite{b4, b5, b19, b20}.
The rate of gradient change throughout training is not uniform either.
Gradients computed in early stages tend to be highly volatile change drastically as a model adjusts its parameters towards the minima aggressively \cite{b21, b1}.
Thus, gradients are large initially and get smaller as a model converges and eventually saturates.
Additionally, there are critical regions in training where gradients are extremely sensitive \cite{b22, b1, b23, b14}.
This volatility is influenced by training hyperparameters like step-size or learning-rate schedule, gradient clipping, SGD variant (e.g., Nesterov, Adam, Adagrad, etc.).
First-order gradients or \emph{variance} approximates second-order information and thus helps detect critical regions and measure the statistical efficiency in data-parallel training \cite{b23, b24, b14}.

\subsection{\textbf{Gradient Compression to Mitigate Communication Cost}}\label{sub:gradcompress}

Lossy gradient compression attenuates communication cost by reducing the volume of gradients exchanged over the network.
Sparsification methods like layerwise Top\textit{k}, or LWTop\textit{k} \cite{b25} and MSTop\textit{k} \cite{b26} retrieve the largest k\% gradients and transmit their respective values and indices, while setting every other tensor element to 0.
The effective CR in this case is thus $k/100$.
Other sparsifiers like DGC \cite{b47} and SIDCo \cite{b48} compress by modeling gradients over some distribution and estimating threshold for a target compression ratio.
Quantization techniques like signSGD \cite{b27} and TernGrad \cite{b28} limit or quantize the gradients to a fixed subset of values, while methods like 1-bit SGD \cite{b29} reduce the bit-width of single-precision gradients.
Low-rank approximators like PowerSGD \cite{b30} decompose $\mathbb{R}^{p \times q}$ dimensional gradients into two lower $r$-rank matrices with dimensions $\mathbb{R}^{p \times r}$ and $\mathbb{R}^{r \times q}$.
Additionally, lossy gradient compression methods rely on \emph{error-feedback} to avoid losing significant updates when using a high degree of compression \cite{b25, b26}.
In Eqn. (\ref{eqn:errorfeedback}), gradients from step $(i-1)$ that are ineligible for communication after applying compression are \emph{not} discarded, but appended to gradients computed in the next iteration, $g_{o}^{(i)}$.
A compression operator $\mathcal{C}$ thus reduces the error-fed gradients $g_{e}^{(i)}$ to $g_{c}^{(i)}$ and updates residual gradients accordingly.

\begin{subequations}
	\begin{gather}
		g_{e}^{(i)} = g_{o}^{(i)} \; + \; \mathtt{residual}^{(i-1)} \\
		g_{c}^{(i)} = \mathcal{C}(g_{e}^{(i)}) \;\;\; \text{and} \;\;\; \mathtt{residual}^{(i)} = g_{e}^{(i)} - g_{c}^{(i)}
	\end{gather}
	\label{eqn:errorfeedback}
\end{subequations}

\vspace{0.12cm}
\subsubsection{Parallel Efficiency in Gradient Compression}\label{subsub:paralleleffcompress}

Although it reduces the synchronization cost and speeds up training, compression itself is \emph{not a zero-cost operation}.
Parsing and compressing gradients before transmission, and deparsing them before applying SGD-update incurs an additional, non-negligible overhead.
With gradients compressed to CR $c$, the per-step time for distributed training in Eqn. (\ref{eqn:itrtime}) reduces to Eqn. (\ref{eqn:compressitrtime}).

\begin{equation}
	t_{step}^{(c)} = t_{compute} + t_{sync}^{(c)} + t_{IO} + t_{comp-decomp}^{(c)}
	\label{eqn:compressitrtime}
\end{equation}

\vspace{0.1cm}
The computation time in Eqn. (\ref{eqn:itrtime}) and (\ref{eqn:compressitrtime}) irrespective of compression since it depends on batch-size and model used.
However, only a fraction of the total updates are sent instead of the entire gradient tensor so as to reduce synchronization time from $t_{sync}$ in DenseSGD to $t_{sync}^{(c)}$.
For compression to be viable, its cumulative compression and communication overhead needs to be lower than the synchronization cost of DenseSGD: $\: t_{sync}^{(c)} + t_{comp-decomp}^{(c)} < t_{sync}$.
Interleaving gradient computation (on subsequent layers) and communication (on current layer) may not be efficient due to contention between them for the same compute resources.

\vspace{0.1cm}
Collective operation also influences the overall speedup in distributed training.
In DenseSGD, gradients are reduced either via PS or AR (as described in \S \ref{subsub:alphabetacost}).
PS and Ring/Tree-AR have low-latency and low-bandwidth cost respectively.
However, the resulting gradients post-compression operation are not compatible with PS or AR-like reductions.
Allgather (AG) is typically used to synchronize compressed updates by transmitting select values and its indexes from each worker, since an index chosen on one worker may not have necessarily been chosen for communication on a different worker.
A gradient tensor of size $G$ compressed to top-$k\%$ values requires transmitting $\left \lceil{2kG/100}\right \rceil$ data-points (i.e., $\left \lceil{kG/100}\right \rceil$ values and $\left \lceil{kG/100}\right \rceil$ indices).
Although some compressors like Random\textit{k} and PowerSGD are allreduce-friendly, the former has a poor convergence quality while the latter has a considerable compression overhead and achieves lower compression rate compared to others \cite{b30}.

\vspace{0.25cm}
\subsubsection{Latency and Bandwidth Variability}\label{subsub:latencybwvariation}

Variable and unpredictable networks are prevalent on edge, cloud and HPC due to a variety of reasons.
Different classes of VMs have varying networking tiers available to them in terms of ingress/egress bandwidth \cite{b37, b38}.
In mobile computing and IoT, edge devices tend to have slower networking compared to fog nodes/servers.
Network heterogeneity can also arise from the topology (e.g., star, mesh, ring, tree, etc.), different links (ethernet, infiniband) and physical mediums (e.g., fiber, copper).
Both latency and bandwidth can vary even over the same link with time due to a variety of factors.
Network bandwidth, which measures the data-transfer rate of a connection can fluctuate over time because:
\begin{itemize}
	\item \emph{Network Congestion:} The effective bandwidth can decrease over high network traffic, increasing data-transfer time on account of saturated links.
	\item \emph{Quality of Service (QoS) prioritization:} Jobs deployed on a shared cluster may have different QoS priorities associated with them, such that high-priority jobs are allocated a bigger chunk of the total bandwidth.
	As a result, low-priority jobs suffer from high-communication cost due to lower available bandwidth.
	\item \emph{Resource Sharing/Contention:} Even without explicit prioritization, concurrent jobs in a shared cluster share finite networking resources, resulting in lower-bandwidth available per-job.
	\item \emph{Network Scheduling:} Switches and routers may use different scheduling algorithms to allocate bandwidth for different traffic flows, which influences how networking resources are shared among applications.
\end{itemize}

The communication delay over a network, or its latency, can fluctuate for jobs running on the edge or even data-centers.
Local area network (LAN) has very low latency, ranging from micro to milliseconds, while inter-rack latency tends to be in the order of tens of milliseconds, while inter-node latency and across data-centers is even worse (see Fig. (\ref{interintranode})).
Network load and congestion can also cause latency-surge under high traffic flows due to queuing delays as multiple jobs perform simultaneous data transfers.
Further, network routing that decides the data-path of a packet also influences its latency.
Although routing protocols dictate the optimal path for transmission, different routes can result in varying latency due to network congestion and physical distance.
As a result, fluctuating latency and bandwidth over a cluster of nodes can affect a job's completion time.

\begin{table}[t]
	\caption{Top\textit{k} Compression + Communication Cost for CRs 0.1 and 0.001 via Allgather (AG) vs. Ring-AR on uncompressed data for different $\alpha$ (ms) and $1/\beta$ (Gbps) on tensors with 100 Million and 1 Billion parameters.}
	\centering
	\begin{tabular}{|c|c|c|c|c|}
		\hline
		\multicolumn{2}{|c|}{\bfseries Configuration} &
      	\multicolumn{3}{c|}{\bfseries Time cost (ms)} \\
      	\hline
      	\bfseries Tensor size & \bfseries ($\alpha, 1/\beta$) & \bfseries AG 0.1 & \bfseries AG 0.001 & \bfseries Ring-AR \\
		\hline
		\multirow{6}{*}{$10^{8}$} & (10, 10) & 525 & 70 & 716 \\		
		\cline{2-5}
		\cline{2-5}
		& (10, 5) & 976 & 74 & 1271 \\
		\cline{2-5}
		& (10, 1) & 4568 & 111 & 5773 \\
		\cline{2-5}
		& (100, 10) & 798 & 340 & 1975 \\
		\cline{2-5}
		& (100, 5) & 1248 & 345 & 2530 \\
		\cline{2-5}
		& (100, 1) & 4830 & 380 & 7028 \\
		\hline
		\multirow{6}{*}{$10^{9}$} & (10, 10) & 5010 & 482 & 5774 \\		
		\cline{2-5}
		& (10, 5) & 9507 & 534 & 11380 \\
		\cline{2-5}
		& (10, 1) & 45355 & 898 & 56190 \\
		\cline{2-5}
		& (100, 10) & 5280 & 745 & 7024 \\
		\cline{2-5}
		& (100, 5) & 9805 & 791 & 12621 \\
		\cline{2-5}
		&(100, 1) & 45645 & 1154 & 57442 \\
		\hline
	\end{tabular}
	\label{table:paralleleffagring}
\end{table}

\vspace{0.1cm}
On 8 inter-node V100s connected via 40Gbps network, we vary $\alpha$ (latency in ms) and $1/\beta$ (bandwidth in Gbps) using the linux \texttt{traffic control} or `\texttt{tc}' module \cite{b39} and measure the time to synchronize tensors with 100 million and 1 billion parameters represented as single-precision floats.
Latency is controlled via \texttt{netem}' (network emulation) qdisc (queuing discipline), while bandwidth is adjusted via `\texttt{htb}' (hierarchical token bucket) qdisc.
We compare the time taken to reduce the original, uncompressed tensors using allreduce (Ring-AR) collective vs. time to communicate tensors compressed to different CRs using allgather (AG) with different network configurations.
For e.g., 1B parameter tensor using CR 0.001 with AG exchanges 2 million data-points, while Ring-AR communication all 1 billion values.
For AG, Table (\ref{table:paralleleffagring}) shows the cumulative time, i.e., sum of compression and communication costs.
Although AG at say, CR $c$ performs better than Ring-AR for the same ($\alpha$, $\beta$) due to smaller message-size, Ring-AR does \emph{not} take $(1/c) \times$ more time to communicate.
This is because the communication cost of different collectives varies with message-size, cluster-size and network parameters.
Ring-AR is bandwidth-optimal but vulnerable to high-latency, as also seen from the wider gap between AG and Ring-AR in 100ms latency case.
Synchronization overhead in AG increases considerably on low-bandwidth networks even when latency stays the same.
These results corroborate the $\alpha$-$\beta$ communication cost model described in Table (\ref{table:comcostops}).


\vspace{0.25cm}
\subsubsection{Statistical Efficiency in Gradient Compression}\label{subsub:stateffcompress}

\begin{table}
	\caption{Step-time, Accuracy and Diff. (w.r.t. DenseSGD) on a 4ms, 20Gbps network.
	DenseSGD uses Ring-AR, LWTop\textit{k}/MSTop\textit{k} use AG.}
	\centering
	\begin{tabular}{|c|c|c|c|c|}
		\hline
		\bfseries Model & \bfseries Method & \bfseries $t_{step}$ (ms) & \bfseries Acc. & \bfseries Diff. \\
		\hline
		\multirow{7}{*}{ResNet18} & DenseSGD & 98.7 & 90.8\% & 0.0\% \\
		\cline{2-5}
		& LWTop\textit{k} 0.1 & 62 & 90.15\% & -0.65\% \\
		\cline{2-5}
		& LWTop\textit{k} 0.01 & 38.4 & 88.49\% & -2.31\% \\
		\cline{2-5}
		& LWTop\textit{k} 0.001 & 36.8 & 87.2\% & -3.6\% \\
		\cline{2-5}
		& MSTop\textit{k} 0.1 & 83.22 & 90.59\% & -0.21\% \\
		\cline{2-5}
		& MSTop\textit{k} 0.01 & 60.22 & 89.83\% & -0.97\% \\
		\cline{2-5}
		& MSTop\textit{k} 0.001 & 58 & 88.69\% & -2.11\% \\
		\hline
		\multirow{7}{*}{ResNet50} & DenseSGD & 152.5 & 98.75\% & 0.0\% \\
		\cline{2-5}
		& LWTop\textit{k} 0.1 & 130.24 & 98.65\% & -0.1\% \\
		\cline{2-5}
		& LWTop\textit{k} 0.01 & 79.4 & 98.3\% & -0.45\% \\
		\cline{2-5}
		& LWTop\textit{k} 0.001 & 72.7 & 96.55\% & -2.2\% \\
		\cline{2-5}
		& MSTop\textit{k} 0.1 & 124.35 & 98.64\% & -0.11\% \\
		\cline{2-5}
		& MSTop\textit{k} 0.01 & 72.63 & 98.49\% & -0.26\% \\
		\cline{2-5}
		& MSTop\textit{k} 0.001 & 67.33 & 97.73\% & -1.02\% \\
		\hline
		\multirow{7}{*}{AlexNet} & DenseSGD & 230.6 & 82.41\% & 0.0\% \\
		\cline{2-5}
		& LWTop\textit{k} 0.1 & 156.8 & 81.68\% & -0.71\% \\
		\cline{2-5}
		& LWTop\textit{k} 0.01 & 33.45 & 81.45\% & -0.96\% \\
		\cline{2-5}
		& LWTop\textit{k} 0.001 & 21.3 & 78.9\% & -3.51\% \\
		\cline{2-5}
		& MSTop\textit{k} 0.1 & 214.5 & 81.8\% & -0.61\% \\
		\cline{2-5}
		& MSTop\textit{k} 0.01 & 91.4 & 81.7\% & -0.71\% \\
		\cline{2-5}
		& MSTop\textit{k} 0.001 & 79 & 78.92\% & -3.49\% \\
		\hline
		\multirow{7}{*}{ViT} & DenseSGD & 475 & 80.4\% & 0.0\% \\
		\cline{2-5}
		& LWTop\textit{k} 0.1 & 362.4 & 79\% & -1.4\% \\
		\cline{2-5}
		& LWTop\textit{k} 0.01 & 94.64 & 78.75\% & -1.65\% \\
		\cline{2-5}
		& LWTop\textit{k} 0.001 & 67.7 & 78.25\% & -2.15\% \\
		\cline{2-5}
		& MSTop\textit{k} 0.1 & 543.6 & 80.47\% & +0.07\% \\
		\cline{2-5}
		& MSTop\textit{k} 0.01 & 275.24 & 79.87\% & -0.53\% \\
		\cline{2-5}
		& MSTop\textit{k} 0.001 & 248.8 & 79.76\% & -0.64\% \\
		\hline
	\end{tabular}
	\label{table:lwmstopkconvergence}
\end{table}

As explained in \S\ref{sub:statefficiency}, distributed training has a statistical efficiency aspect associated with it due to gradient \emph{variance} or \emph{noise}.
This is further exacerbated in lossy compression where model weights are updated with an even less accurate gradient representation \cite{b1}.
To see the impact of varying degrees of compression on convergence, we train DNNs with \emph{LWTop\textit{k}} \cite{b25} and \emph{MSTop\textit{k}} \cite{b26} at CRs [0.1, 0.01, 0.001].
We compare final model accuracy with that of DenseSGD and observe differences in their convergence quality.
Layerwise or LWTop\textit{k} retrieves top-$k$\% gradient values and indexes on a layer-by-layer basis, while MSTop\textit{k} approximates top-\textit{k} on the entire gradient tensor (i.e., not on layerwise basis) via multi-sampling and uses binary search to find the threshold corresponding to target CR. 
Threshold estimation in MSTop\textit{k} involves multiple trials or searches for a configurable number of rounds to predict an accurate threshold; we use 25 rounds in our evaluation.
Training routine and hyperparameters are described in \S\ref{subsub:trainschedule}.

 \begin{figure}
	\subfloat[LWTop\textit{k} Compression]{\includegraphics[width=0.23\textwidth]{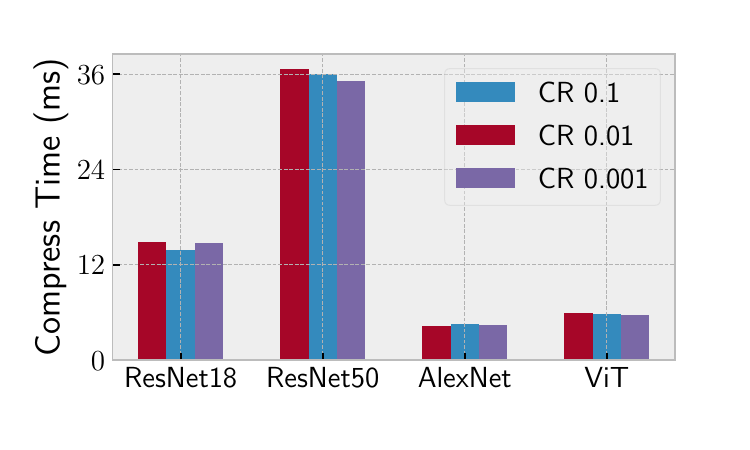}
	\label{lwtopkoverhead}}
	\hfill
	\subfloat[MSTop\textit{k} Compression]{\includegraphics[width=0.23\textwidth]{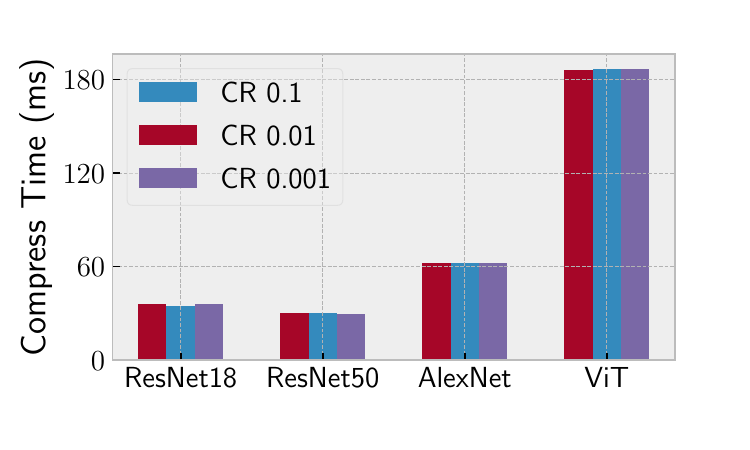}
	\label{mstopkoverhead}}
	\caption{Compression overhead of different techniques. For the same CR, MSTop\textit{k} has higher compression cost due to multi-round threshold estimation.}	
	\label{fig:lwmstopcompoverhead}
\end{figure}

\vspace{0.11cm}
Table (\ref{table:lwmstopkconvergence}) tabulates the step-time, test accuracy and convergence difference (w.r.t. DenseSGD) while training on the same cluster of 8 inter-node GPUs when the latency and bandwidth are limited to 4ms and 20Gbps via \texttt{tc}.
Step-time includes gradient computation, IO overhead, compression and communication cost for different CRs using AG-collective, while DenseSGD has no compression cost but transmits entire gradients via AR-collective.
Further, the step-time of MSTop\textit{k} is higher than LWTop\textit{k} irrespective of the model or CR.
This is because the former has higher compression cost on account of multiple rounds needed to compute the threshold, as also seen in Fig. (\ref{fig:lwmstopcompoverhead}).
In some cases, like ViT at CR 0.1, MSTop\textit{k} has higher iteration time than even DenseSGD because of high compression overhead.
However, MSTop\textit{k} achieves similar or \emph{even better} accuracy than LSTop\textit{k} since its estimates threshold over the entire model, while the latter compresses on a layerwise basis.
Thus, in LWTop\textit{k}, models with non-uniform layers and skewed gradients may lose more critical updates since the fraction of gradients selected from each layer is fixed (in proportion to the layer's size).
At the same time, more significant updates from larger layers may not even be eligible for communication if most of the critical updates are cluttered in only a few layers.

\vspace{0.08cm}
ResNet18 converges to 90.8\% accuracy with DenseSGD, while LSTop\textit{k} achieves 88.49\% and 87.2\% accuracy for CRs 0.01 and 0.001 respectively.
MSTop\textit{k} on ResNet18 attains 89.83\% and 88.69\% for similar CRs.
We observe a similar pattern for all other DNNs; as more compression is applied and CR is lowered, model convergence degrades as well.
This happens because we lose considerably more information at smaller CRs.
GraVAC \cite{b1} defines a heuristic called \emph{Compression gain} to measure this loss by comparing prior and post-compression gradients.
It compares the variance between error-fed and compressed gradients (i.e., $g_{e}^{(i)}$ and $g_{c}^{(i)}$ from Eqn. (\ref{eqn:errorfeedback})) at step $i$ as: $\mathbf{\mathtt{E}[||g_{c}^{(i)}||^{2}]\: / \: \mathtt{E}[||g_{e}^{(i)}||^{2}]}$.

\begin{figure}[t]
	\centering 
	{\includegraphics[width=0.35\textwidth]{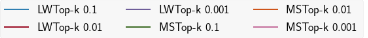}\vspace*{-0.33cm}}
	\hspace{0.01cm}
	\subfloat[ResNet18]{\includegraphics[width=0.23\textwidth]{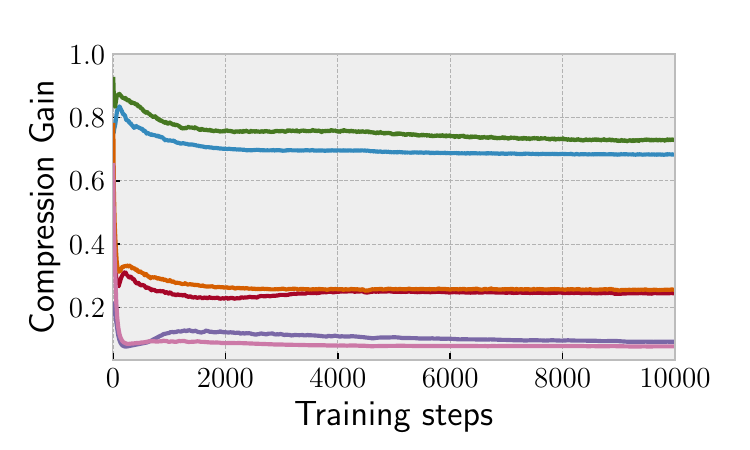}
	\label{resnet18lwmstopk}}
	\subfloat[ResNet50]{\includegraphics[width=0.23\textwidth]{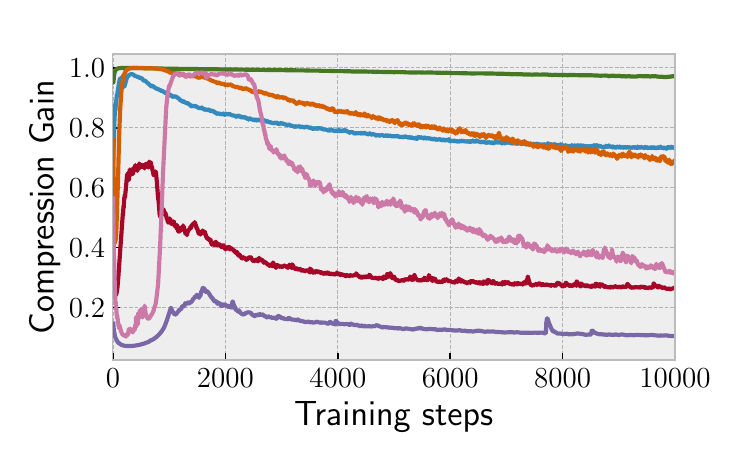}
	\label{resnet50lwmstopk}}
	\hspace{0.01cm}
	\subfloat[AlexNet]{\includegraphics[width=0.23\textwidth]{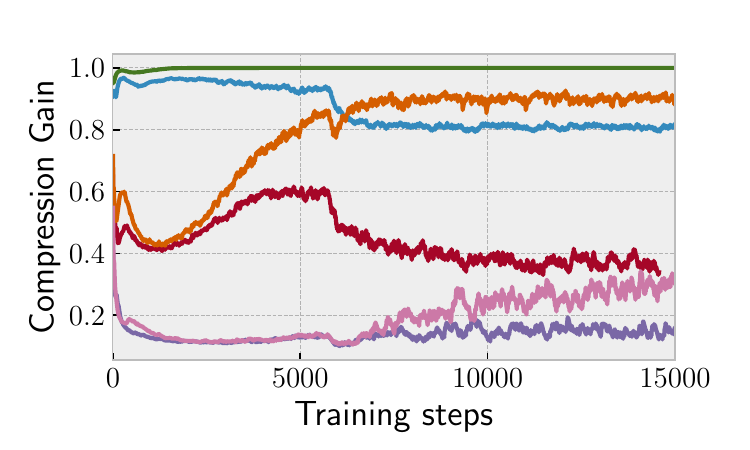}
	\label{alexnetlwmstopk}}
	\subfloat[ViT]{\includegraphics[width=0.23\textwidth]{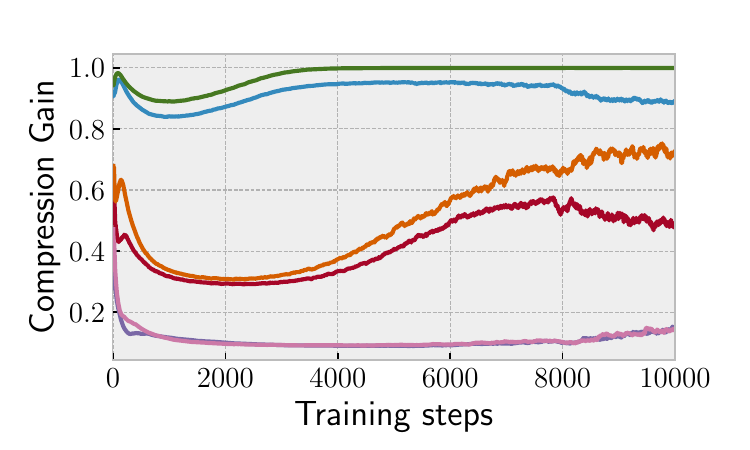}
	\label{vitlwmstopk}}
	\caption{Compression gain measures statistical efficiency in lossy compressors.}
	\label{fig:lwmstopkbaseline}
\end{figure}

\vspace{0.1cm}
The overhead of computing compression gain is low as it only needs first-order gradients already computed in backpropagation. 
We plot gains of LWTop\textit{k} and MSTop\textit{k} at various CRs in Fig. (\ref{fig:lwmstopkbaseline}) and see that gain tends to be smaller for lower CRs as more gradient information is lost there.
Gain also changes in early training stages and certain critical regimes \cite{b1, b21, b22, b23} (like step-size decay in our case) before it eventually saturates.
Additionally, we see a correlation between compression gain and model convergence for each compressor and CR configuration.
From Table (\ref{table:lwmstopkconvergence}), ResNet50 at CR 0.001 achieves 96.55\% accuracy with LWTop\textit{k} and a higher 97.73\% with MSTop\textit{k}.
At the same time, MSTop\textit{k} has a higher compression gain than LWTop\textit{k}, seen in Fig. (\ref{resnet50lwmstopk}).
Further, the generalization performance of different compressors at different CRs can also be inferred by their compression gains.
For e.g., AlexNet with  LWTop\textit{k} at CR 0.1 and MSTop\textit{k} with CR 0.01 have roughly the same 81.68\% and 81.7\% accuracy, while their gains also saturate to the same level (Fig. (\ref{alexnetlwmstopk})).
Thus, we see how compression gain works as a practical and effective heuristic to evaluate the efficacy of a compressor and measure statistical efficiency in lossy gradient compression.
	\section{AR-Top\textit{k} Compression}\label{sec:artopkcompress}

In this paper, we propose a new compression technique compatible with AR-collective called \emph{AR-Top\textit{k}} along with its two variants, followed by evaluation and communication cost analysis of the same.
However, as we will see, aggregating updates via AR collective may not necessarily be better than AG, depending on network and distributed training parameters.
We thus propose a flexible compression-communication scheme that utilizes the faster collective based on the $\alpha$-$\beta$ communication model.
To determine the ideal CR and collective to use (either AG, AR-Top\textit{k} with ring or tree-AR) that best balances parallel and statistical efficiency when training with lossy compression, we model gradient compression as a multi-objective optimization (MOO) problem.

\subsection{\textbf{Allreduce-friendly Top\textit{k} Compression (AR-Top\textit{k})}}\label{sub:allreducetopk}

We implement Top-\textit{k} sorting using max-heap and retrieve top $k\%$ values over fused gradient tensors, i.e., compression is applied over the entire model and not just layerwise (as with LWTop\textit{k}).
Compared to MSTop\textit{k} which is based on threshold estimation, the compression overhead of max-heap based sorting is much lower on GPUs.
Between AG and AR, the collective that incurs the least communication overhead in distributed training depends on the message-size to transmit, cluster-size, target CR and network-specific parameters like latency and bandwidth.
The message size is doubled with AG as both values and indices are exchanged, while AR is not out-of-the-box compatible with lossy compression.
AR-Top\textit{k} takes the compressed gradient values and indexes from Top-\textit{k} sorting and initiates a \texttt{Broadcast} from a worker to disperse its corresponding indices, followed by \texttt{AllReduce} to aggregate values on indices that were broadcasted in the first step.
The AR call in second step can use either ring or tree-based reduction (covered in \S\ref{sub:commcostartopk}).

\vspace{0.1cm}
\subsection{\textbf{AR-Top\textit{k} Worker Selection Mechanism}}\label{sub:workerselecrartopk}

\begin{algorithm}[t]
	\label{alg:artopkalgo}
	\DontPrintSemicolon
	\SetKwProg{Pn}{function}{:}{\KwRet}
	\SetKwFunction{train}{train}
	\SetKwFunction{var}{var}
	\SetKwFunction{bcast}{Broadcast}
	\SetKwFunction{allgather}{AllGather}
	\SetKwFunction{allreduce}{AllReduce}
	\SetKwFunction{topk}{Top\textit{k}}
	\SetKwFunction{vartopk}{VAR-Top\textit{k}}
	\SetKwFunction{startopk}{STAR-Top\textit{k}}
	\caption{AR-Top\textit{k} Compression}
	\textbf{Input:} CR $c$, step-size $\eta$, cluster-size $N$, worker-rank $r$\\
	\Pn{\train{}}{
		$\text{residual}_{(0, r)} \:=\: 0$\\
		\For {$\text{i} = 1,2,...$}{
			$\text{G}_{(i, r)} = \nabla\mathcal{F}(x_{(i, r)}, w_{(i, r)}) + \text{residual}_{(i-1, r)}$\\
			$\text{g}_{(i, r)}$, $\text{ix}_{(i, r)}$ = \topk{$\text{G}_{(i, r)}, \;c$}\\
			\If{\startopk}{
				$\tilde{r}_{(i)} = i \;\: \% \;\:N$\\
			\BlankLine
			}\ElseIf{\vartopk}{
				\var[$N$] = 0\\
				\var[$r$] = $||g_{(i, r)}||^{2}$\\
				\var = \allgather{\var}\\
				$\tilde{r}_{(i)} = \var.\text{indexOf}(\text{max}(\var))$\\
			}
			\BlankLine
			$\tilde{\text{ix}}_{(i, r)} =\;$\bcast{$\text{ix}_{(i, r)}, src=\tilde{r}_{(i)}$}\\
			$\tilde{\text{g}}_{(i, r)} = \text{G}_{(i, r)}$[$\tilde{\text{ix}}_{(i, r)}$]\\
			$\text{residual}_{(i, r)} = \text{G}_{(i, r)} - \tilde{\text{g}}_{(i, r)}$\\
			$\tilde{\text{g}}_{(i, r)} =\;$\allreduce{$\tilde{\text{g}}_{(i, r)}$}\\
			$w_{(i+1, r)} = w_{(i, t)} - \eta\cdot \tilde{\text{g}}_{(i, r)}$\\
			$\text{i}$++\\
		}
	}
\end{algorithm}

After broadcasting the compressed indices from a worker, gradient values reduced at those positions may not necessarily correspond to top-\textit{k} gradients across all workers.
\textit{Additionally, the natural question that then arises is, `How do we choose which worker gets to broadcast its indices to all other workers in the cluster for aggregating its local top-\textit{k} values?'}
We explore two mechanisms to determine worker selection in AR-Top\textit{k}, one \textit{variance-based} and the other \textit{staleness-based}:

\begin{enumerate}
	\vspace{0.05cm}
	\item \emph{\textbf{Variance-based AR-Topk (VAR-Topk)}} selects worker with largest gradient variance and disperses the indexes of its local top-\textit{k} values, followed by calling AR on those elements.
As discussed in \S\ref{sub:statefficiency} and \S\ref{subsub:stateffcompress}, the significance of updates can be measured by their gradient variance.
Thus, worker with maximum variance sets the tone for calling AR-collective to aggregate values at indices corresponding to its local top-\textit{k} values. 
	\vspace{0.1cm}
	\item \emph{\textbf{Staleness-based AR-Topk (STAR-Topk)}} chooses top-\textit{k} indices from workers in a round-robin fashion, which are then broadcasted to call AR on gradients at those indices across all workers.
In this way, each worker gets to transmit its compressed update once every $N$ iterations (in a cluster of $N$ workers).
While error-feedback effectively works as \emph{delayed} or \emph{stale} updates, STAR-Top\textit{k} further adds to it by adding an implicit staleness of $N$ steps.
\end{enumerate}

\vspace{0.1cm}
Alg. (\ref{alg:artopkalgo}) illustrates how workers are selected in AR-Top\textit{k} compression.
Error-fed gradients $\text{G}_{(i, r)}$ at step `$i$' are compressed to CR `$c$' on each worker `$r$' in the first step.
STAR-Top\textit{k} selects a worker based on the current iteration number and cluster-size (line 8 in Alg. (\ref{alg:artopkalgo})).
VAR-Top\textit{k} involves additional steps/overhead, such that an array of size `$N$' holds the gradient variance of worker `$r$' at index `$r$' (line 11).
An AG call on the array synchronizes variance across all workers (line 12), and the worker with maximum variance is chosen for communication (line 13).
\emph{The overhead of this step is low as the message-exchange size is just `$N$' 32-bit floats.}
Once a worker (denoted by rank $\tilde{r}_{(i)}$) is selected either by STAR or VAR-Top\textit{k}, its corresponding indices are dispersed via broadcast (line 14)
The gradients residing at those indexes on the remaining workers are then selected, residual gradients are updates and finally aggregated via AR-collective (lines 15-17).
Lastly, model parameters are updated and training proceeds to the next iteration (lines 18-19).

\vspace{0.05cm}
\subsection{\textbf{Training Speedup and Convergence in AR-Top\textit{k}}}\label{sub:convergenceartopk}

\begin{table}[t]
	\caption{Convergence and $t_{step}$ (ms) in DenseSGD and STAR/VAR-Top\textit{k}.
	DenseSGD uses Tree-AR on 4ms latency and 20Gbps bandwidth network.}
	\centering
	\begin{tabular}{|c|c|c|c|c|}
	\hline
	\bfseries Model & \bfseries Method & \bfseries $t_{step}$ (ms) & \bfseries Acc. & \bfseries Diff.\\
	\hline
	\multirow{7}{*}{ResNet18} & DenseSGD & 146.21 & 90.79\% & 0.0\% \\
	\cline{2-5}
	& STAR-Top\textit{k} 0.1 & 64.83 & 90.19\% & -0.6\% \\
	\cline{2-5}
	& STAR-Top\textit{k} 0.01 & 49.68 & 89.89\% & -0.9\% \\
	\cline{2-5}
	& STAR-Top\textit{k} 0.001 & 48.17 & 88.81\% & -1.98\% \\
	\cline{2-5}
	& VAR-Top\textit{k} 0.1 & 77.2 & 90.15\% & 0.64\% \\
	\cline{2-5}
	& VAR-Top\textit{k} 0.01 & 62.1 & 89.5\% & -1.29\% \\
	\cline{2-5}
	& VAR-Top\textit{k} 0.001 & 60.5 & 88.7\% & -2.09\% \\
	\hline
	\multirow{7}{*}{ResNet50} & DenseSGD & 294.35 & 98.72\% & 0.0\% \\
	\cline{2-5}
	& STAR-Top\textit{k} 0.1 & 100.8 & 98.68\% & -0.04\% \\
	\cline{2-5}
	& STAR-Top\textit{k} 0.01 & 67.68 & 98.57\% & -0.15\% \\
	\cline{2-5}
	& STAR-Top\textit{k} 0.001 & 64.37 & 98.38\% & -0.34\% \\
	\cline{2-5}
	& VAR-Top\textit{k} 0.1 & 113.4 & 97.65\% & -1.07\% \\
	\cline{2-5}
	& VAR-Top\textit{k} 0.01 & 80 & 98.2\% & -0.52\% \\
	\cline{2-5}
	& VAR-Top\textit{k} 0.001 & 76.5 & 97.33\% & -1.39\% \\
	\hline
	\multirow{7}{*}{AlexNet} & DenseSGD & 614 & 82.4\% & 0.0\% \\
	\cline{2-5}
	& STAR-Top\textit{k} 0.1 & 129.5 & 81.74\% & -0.66\% \\
	\cline{2-5}
	& STAR-Top\textit{k} 0.01 & 50.3 & 80.41\% & -1.99\% \\
	\cline{2-5}
	& STAR-Top\textit{k} 0.001 & 42.38 & 79.57\% & -2.83\% \\
	\cline{2-5}
	& VAR-Top\textit{k} 0.1 & 142 & 83.15\% & +0.75\% \\
	\cline{2-5}
	& VAR-Top\textit{k} 0.01 & 62.6 & 81.78\% & -0.62\% \\
	\cline{2-5}
	& VAR-Top\textit{k} 0.001 & 54.4 & 78.06\% & -4.34\% \\
	\hline
	\multirow{7}{*}{ViT} & Dense-SGD & 1348.5 & 80.42\% & 0.0\% \\
	\cline{2-5}
	& STAR-Top\textit{k} 0.1 & 276.32 & 79.72\% & -0.7\% \\
	\cline{2-5}
	& STAR-Top\textit{k} 0.01 & 104.13 & 79.18\% & -1.24\% \\
	\cline{2-5}
	& STAR-Top\textit{k} 0.001 & 86.91 & 79.23\% & -1.19\% \\
	\cline{2-5}
	& VAR-Top\textit{k} 0.1 & 289.2 & 80.12\% & -0.3\% \\
	\cline{2-5}
	& VAR-Top\textit{k} 0.01 & 117 & 79.2\% & -1.22\% \\
	\cline{2-5}
	& VAR-Top\textit{k} 0.001 & 99.7 & 79.8\% & -0.62\% \\
	\hline
	\end{tabular}
	\label{table:artopkconvergence}
\end{table}

\subsubsection{DNNs and Hyperparameters}\label{subsub:trainschedule}

The following models are deployed with PyTorch verison v1.12.1 \cite{b16}, using NCCL v2.10.3 for communication between 8 inter-node GPUs.
ResNet18 trains on CIFAR100 \cite{b35} for 50 epochs with per-worker batch-size 32, momentum 0.1, weight decay 0.0005, step-size 0.1 and decay factor 0.1 at epoch 15, 30 and 45.
ResNet50 runs on Food101 \cite{b43} dataset for 100 epochs with batch-size 64, momentum 0.1, step-size 0.1 and weight decay 0.0001.
AlexNet trains on Caltech101 \cite{b40} for 50 epochs with batch-size 32, momentum 0.9, weight decay 0.0005, step-size 0.01 and decay factor 0.1 at epoch 25.
ViT runs on Caltech256 \cite{b41} for 50 epochs with batch-size 32, momentum 0.9, weight decay 0.0005, step-size 0.01 and decay rate 0.2 on epoch 40.
We report the top-1 test accuracy and time to complete one training step (in ms).

\vspace{0.2cm}
\subsubsection{Training with STAR-Top\textit{k} and VAR-Top\textit{k} Compression}

 \begin{figure}[t]
	 \subfloat[ViT with CR 0.01]{\includegraphics[width=0.23\textwidth]{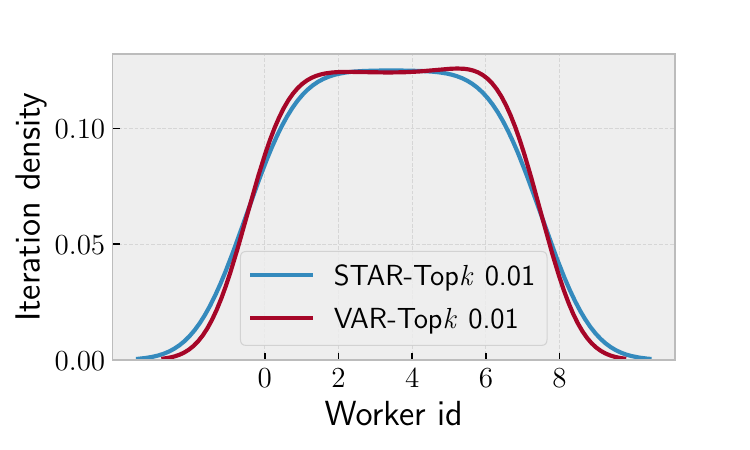}
	\label{fig:starvarvgg}}
	\hfill
	\subfloat[AlexNet with CR 0.001]{\includegraphics[width=0.23\textwidth]{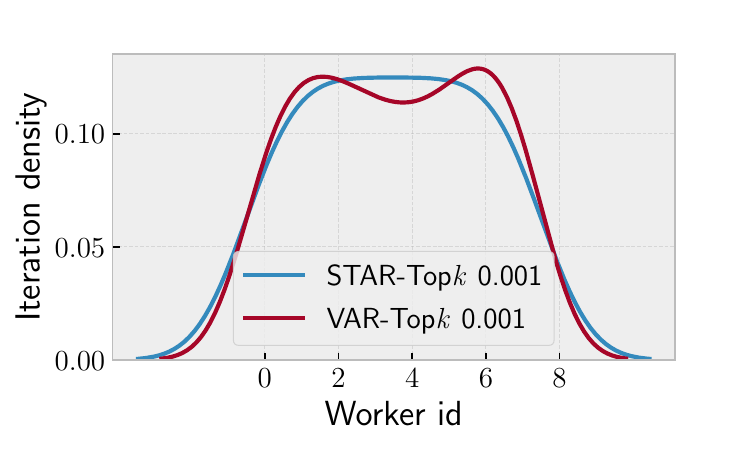}
	\label{fig:starvaralexnet}}
	\caption{Iteration density of the 8 workers (ranked 0-7) that broadcast their top\textit{k} indices with STAR and VAR-Top\textit{k} compression.
	STAR-Top\textit{k} has similar density across all workers due to its round-robin nature, while some workers may have higher density of updates over others in VAR-Top\textit{k}.}
	\label{fig:itrdensityartopk}
\end{figure}

Table (\ref{table:artopkconvergence}) compares the model convergence and per-iteration speedup in STAR and VAR-Top\textit{k} with DenseSGD.
We train on a network limited to 4ms latency and 20Gbps bandwidth via \texttt{tc}, and configure NCCL to use Tree-AR in DenseSGD by setting environment variable \texttt{NCCL\_ALGO} to `tree'.
The general trend is that accuracy decreases as CR is lowered in both STAR and VAR-Top\textit{k}.
For the same CR, STAR-Top\textit{k} performs better than VAR-Top\textit{k} at the current cluster-size and network configuration.
For ResNet50 CR 0.001, STAR-Top\textit{k} had 0.34\% lower accuracy than DenseSGD, while VAR-Top\textit{k} converged to 1.39\% lower accuracy.
To understand the deviation in generalization quality between STAR and VAR-Top\textit{k} even at the same CR, we look at the iteration kernel density estimates (KDE) of each worker that broadcasts its top\textit{k} indices over the entire training, with worker ids ranked from 0 to ($N-1$).
We see similar iteration densities of STAR and VAR-Top\textit{k} in ViT CR 0.01 (Fig. (\ref{fig:starvarvgg})), which also have close accuracy as shown in Table (\ref{table:artopkconvergence}); STAR-Top\textit{k} achieves 79.18\% while VAR-Top\textit{k} saturates to 79.2\%.
With AlexNet at CR 0.001, STAR and VAR-Top\textit{k} converged to 79.57\% and 78.06\% respectively, a considerable 1.51\% difference.
In this case, iteration density of VAR-Top\textit{k} is more skewed and non-uniform compared to STAR-Top\textit{k}, as seen in Fig. (\ref{fig:starvaralexnet}).
Worker ranks 1 and 6 broadcast considerably more updates than others, while rank 4 transmits its top\textit{k} gradients much \emph{less frequently}.
The infrequent updates of rank 4 are thus more stale which in turn degrades model quality.
On the other hand, STAR-Top\textit{k} has a much more even and uniform iteration density due to its round-robin approach.

\vspace{0.1cm}
Additionally, iteration time is higher in VAR-Top\textit{k} over STAR-Top\textit{k}.
This is because the former involves additional steps over the latter, as illustrated in lines (10-13) of Alg. (\ref{alg:artopkalgo}).
VAR-Top\textit{k} makes an additional AG call to collect gradient variance from all workers, exchanging $4N$ bytes of data in the process.
The overhead of this AG call is negligible due to its small message-size.
\emph{Although STAR-Top\textit{k} performs better both in terms of parallel (due to its lower compute overhead) and statistical efficiency (as it achieves better accuracy with its round-robin approach), we conjecture that VAR-Top\textit{k} compression is also a viable approach, especially when training on unbalanced and non-i.i.d. (independent and identically distributed) data or when training over a massive number of devices, as commonly seen in federated learning.
With unbalanced data, collecting updates from workers with large variance prioritizes critical updates.
When cluster-size $N$ is huge, the round-robin updates of STAR-Top\textit{k} can get too stale and degrade model convergence.}

\vspace{0.2cm}
\subsubsection{Performance of AR-Top\textit{k} over LWTop\textit{k} Compression}

\begin{table}
	\caption{Comparing STAR-Top\textit{k}, VAR-Top\textit{k} and LWTop\textit{k} on ResNet18/50, AlexNet and ViT.
	AR-Top\textit{k} variants use Allreduce, LWTop\textit{k} uses Allgather.}
	\centering
	\begin{tabular}{|c|c|c|c|c|c|c|c|}
		\hline
		\multicolumn{2}{|c|}{\bfseries DNN config} & \multicolumn{3}{c|}{\bfseries $t_{step}$ (ms)} & \multicolumn{3}{c|}{\bfseries Accuracy (\%)} \\
      	\hline
      	\bfseries Model & \bfseries CR & \bfseries STAR & \bfseries VAR & \bfseries LW & \bfseries STAR & \bfseries VAR & \bfseries LW \\
      	\hline
      	\multirow{3}{*}{Res.18} & 0.1 & 64.83 & 77.2 & \textbf{62} & \textbf{90.19} & 90.15 & 90.15 \\
      	\cline{2-8}
      	& 0.01 & 49.68 & 62.1 & \textbf{38.4} & \textbf{89.89} & 89.5 & 88.49 \\
      	\cline{2-8}
      	& 0.001 & 48.17 & 60.5 & \textbf{36.8} & \textbf{88.81} & 88.7 & 87.2 \\
      	\hline
      	\multirow{3}{*}{Res.50} & 0.1 & \textbf{100.8} & 113.4 & 130.24 & \textbf{98.68} & 97.65 & 98.65 \\
      	\cline{2-8}
      	& 0.01 & \textbf{67.68} & 80 & 79.4 & \textbf{98.57} & 98.2 & 98.3 \\
      	\cline{2-8}
      	& 0.001 & \textbf{64.37} & 76.5 & 72.7 & \textbf{98.38} & 97.33 & 96.55 \\
      	\hline
      	\multirow{3}{*}{Alex.} & 0.1 & \textbf{129.5} & 142 & 156.8 & 81.74 & \textbf{83.15} & 81.68 \\
      	\cline{2-8}
      	& 0.01 & 50.3 & 62.6 & \textbf{33.45} & 80.41 & \textbf{81.78} & 81.45 \\
      	\cline{2-8}
      	& 0.001 & 42.38 & 54.4 & \textbf{21.3} & \textbf{79.57} & 78.06 & 78.9 \\
      	\hline
      	\multirow{3}{*}{ViT} & 0.1 & \textbf{276.32} & 289.2 & 362.4 & 79.72 & \textbf{80.12} & 79 \\
      	\cline{2-8}
      	& 0.01 & 104.13 & 117 & \textbf{94.64} & 79.18 & \textbf{79.2} & 78.75 \\
      	\cline{2-8}
      	& 0.001 & 86.91 & 99.7 & \textbf{67.7} & 79.23 & \textbf{79.8} & 78.25 \\
      	\hline
	\end{tabular}
	\label{tab:ARtopvsLWvsMS}
\end{table}

In this section, we directly compare AR-Top\textit{k} (both variants) with LWTop\textit{k} in Table (\ref{tab:ARtopvsLWvsMS}) where the CR is fixed throughout training.
STAR/VAR-Top\textit{k} use AR collective communication, while LWTop\textit{k} uses AG.
The network latency is limited to 4ms, while bandwidth is capped at 20Gbps.
We observe that STAR-Top\textit{k} has a lower iteration time than LWTop\textit{k} in some cases, and higher in others.
For e.g., STAR-Top\textit{k} is faster than LWTop\textit{k} at all CRs of ResNet50, and CR 0.1 in AlexNet and ViT.
Inversely, LWTop\textit{k} is faster at all CRs in ResNet18, and CRs (0.01, 0.001) in AlexNet and ViT.
VAR-Top\textit{k} has higher iteration cost across all models due to its additional variance compute operation.
Thus, AG works better than AR in some cases, and vice versa, determined by network topology, latency, bandwidth, CR, model and cluster-size.
AR-Top\textit{k} applies tensor fusion prior compression, i.e., we compress gradients as a whole across all layers while LWTop\textit{k} does so on a layer-by-layer basis.
This appeal of the former approach can be seen from model accuracy in Table (\ref{tab:ARtopvsLWvsMS}); STAR-Top\textit{k} achieves higher accuracy across all CRs of ResNet18 and ResNet50, while VAR-Top\textit{k} attains higher accuracy for majority of the CRs in AlexNet and ViT.

\subsection{\textbf{Communication Cost Analysis of AR-Top\textit{k}}}\label{sub:commcostartopk}

Communication cost of AG to exchange $2Mc$ bytes of gradient values and indices is: $\alpha\log(N) + 2Mc\beta (N-1)$, where the network latency is denoted by $\alpha$ and $1/\beta$ is the available bandwidth to transmit $M$ bytes of gradient information that are compressed to CR $c$.
With AR-Top\textit{k}, communication is a two-step process: one Broadcast to distribute top-\textit{k} indexes from a single worker, followed by AR to reduce gradient values at those specific indexes.
AR-Top\textit{k} can either use Ring or Tree-AR during reduction phase, which have different costs associated with each of them (see Table (\ref{table:comcostops})).
Thus, communication cost of AR-Top\textit{k} depends on the specific flavor of AR algorithm used, shown in Eqn. (\ref{eqn:artopkcommRing}) and (\ref{eqn:artopkcommTree}) for AR-Top\textit{k} with Ring and Tree reduction respectively.

\begin{subequations}
	\begin{equation}
		t_{ARTring}^{(c)} = \alpha[\:2(N-1) \:+\: \log (N)\:] + Mc\beta [\;2\dfrac{(N-1)}{N} \:+\: \log (N)\;]
		\label{eqn:artopkcommRing}
	\end{equation}
	\begin{equation}
		t_{ARTtree}^{(c)} = 3\alpha \log (N) + 3Mc\beta \log (N)
		\label{eqn:artopkcommTree}
	\end{equation}
	\label{eqn:comcostartopk}
\end{subequations}

\begin{table}
	\caption{Communication cost of varying bandwidths with fixed latency of 1ms for AR-Top\textit{k} (ART) with Ring and Tree-AR, and with AG-collective.}
	\centering
	\begin{tabular}{|c|c|c|c|c|c|}
		\hline
		\multicolumn{2}{|c|}{\bfseries Configuration} &
      	\multicolumn{4}{c|}{\bfseries Communication Time (ms)} \\
      	\hline
		\bfseries Model & \bfseries ($\alpha,1/\beta$) & \bfseries CR & \bfseries AG & \bfseries ART-Ring & \bfseries ART-Tree \\
		\hline
		\multirow{9}{*}{ResNet18} & \multirow{3}{*}{(1,10)} & 0.1 & 54 & \textbf{35} & 43.2 \\
		\cline{3-6}
		& & 0.01 & \textbf{7.66} & 18.1 & 12.2 \\
		\cline{3-6}
		& & 0.001 & \textbf{3.28} & 16.7 & 9 \\
		\cline{2-6}
		& \multirow{3}{*}{(1,5)} & 0.1 & 107.76 & \textbf{52.5} & 76.3 \\
		\cline{3-6}
		& & 0.01 & \textbf{13.83} & 20.8 & 16.1 \\
		\cline{3-6}
		& & 0.001 & \textbf{4.25} & 17.9 & 10.1 \\
		\cline{2-6}
		& \multirow{3}{*}{(1,1)} & 0.1 & 526.3 & \textbf{194.7} & 345.6 \\
		\cline{3-6}
		& & 0.01 & 51.93 & \textbf{34.1} & 41.9 \\
		\cline{3-6}
		& & 0.001 & \textbf{8.86} & 19.5 & 12.8 \\
		\hline
		\multirow{9}{*}{ResNet50} & \multirow{3}{*}{(1,10)} & 0.1 & 115.1 & \textbf{52.9} & 83.4 \\
		\cline{3-6}
		& & 0.01 & \textbf{14.35} & 20.3 & 15.9 \\
		\cline{3-6}
		& & 0.001 & \textbf{4.65} & 18.1 & 10 \\
		\cline{2-6}
		& \multirow{3}{*}{(1,5)} & 0.1 & 232 & \textbf{94.7} & 156.2 \\
		\cline{3-6}
		& & 0.01 & 28.1 & 26.1 & \textbf{24.2} \\
		\cline{3-6}
		& & 0.001 & \textbf{5.3} & 17.8 & 10.5 \\
		\cline{2-6}
		& \multirow{2}{*}{(1,1)} & 0.1 & 1148 & \textbf{405.5} & 745 \\
		\cline{3-6}
		& & 0.01 & 126.5 & \textbf{58.8} & 83.7 \\
		\cline{3-6}
		& & 0.001 & \textbf{14.35} & 21 & 16.1 \\
		\hline
		\multirow{9}{*}{AlexNet} & \multirow{3}{*}{(1,10)} & 0.1 & 271.8 & \textbf{106.8} & 180.4 \\
		\cline{3-6}
		& & 0.01 & 32.73 & \textbf{25.2} & 25.8 \\
		\cline{3-6}
		& & 0.001 & \textbf{6} & 18.6 & 11.1 \\
		\cline{2-6}
		& \multirow{3}{*}{(1,5)} & 0.1 & 544.5 & \textbf{200.4} & 354.8 \\
		\cline{3-6}
		& & 0.01 & 61.75 & \textbf{34.8} & 42.6 \\
		\cline{3-6}
		& & 0.001 & \textbf{8.92} & 19.3 & 13.1 \\
		\cline{2-6}
		& \multirow{3}{*}{(1,1)} & 0.1 & 2718.7 & \textbf{964.4} & 1778 \\
		\cline{3-6}
		& & 0.01 & 282.7 & \textbf{111.8} & 186.8 \\
		\cline{3-6}
		& & 0.001 & 31.33 & \textbf{27} & 27.3 \\
		\hline
		\multirow{9}{*}{ViT} & \multirow{3}{*}{(1,10)} & 0.1 & 592.77 & \textbf{238.6} & 401.2 \\
		\cline{3-6}
		& & 0.01 & 68.48 & \textbf{36.2} & 46.2 \\
		\cline{3-6}
		& & 0.001 & \textbf{9.15} & 19.2 & 12.9 \\
		\cline{2-6}
		& \multirow{3}{*}{(1,5)} & 0.1 & 1206 & \textbf{424.3} & 779.1 \\
		\cline{3-6}
		& & 0.01 & 127.45 & \textbf{58} & 86.2 \\
		\cline{3-6}
		& & 0.001 & \textbf{15.3} & 21.4 & 16.9 \\
		\cline{2-6}
		& \multirow{3}{*}{(1,1)} & 0.1 & 5973 & \textbf{2047} & 3852 \\
		\cline{3-6}
		& & 0.01 & 601.8 & \textbf{222.8} & 385.2 \\
		\cline{3-6}
		& & 0.001 & 59.68 & \textbf{36.7} & 44.4 \\
		\hline
	\end{tabular}
	\label{table:simcommartopkAG}
\end{table}

We limit $\alpha$ to 1 ms and vary $1/\beta$ as 10, 5 and 1Gbps to measure communication time between AG, AR-Top\textit{k} (ART) with Ring and Tree-AR for various DNNs and CRs on a cluster of 8 inter-node V100s.
We fuse tensors into 64MB buckets with PyTorch gradient-bucketing and list the communication cost of each collective in Table (\ref{table:simcommartopkAG}).
At a moderately-high 10Gbps bandwidth and CR 0.1, ART-Ring has the least communication overhead across all DNNs.
On the other hand, AG-collective has the least communication cost at very low CRs like 0.001 and high-to-moderate bandwidth (like 10 and 5Gbps).
In low-bandwidth settings, AR-Top\textit{k} had the advantage over AG, since the AR call made in the former is bandwidth-optimal.
ART-Ring was the optimal collective in low-latency, high-bandwidth (10Gbps) and high CR (0.1) scenarios as well.
\emph{Thus, the optimal collective operation to use in distributed training with gradient compression involves a large trade-off space and depends on network latency, bandwidth, model-size, cluster-size and CR.
AG generally performs better when model size is relatively small (like ResNet18) with moderately-high network bandwidth or very low CRs.
In contrast, AR-Top\textit{k} with ring or tree performs better larger models in low-bandwidth cases, as well as high CRs on high-bandwidth networks.}


 \begin{figure}[t]
	 \subfloat[AllGather communication]{\includegraphics[width=0.23\textwidth]{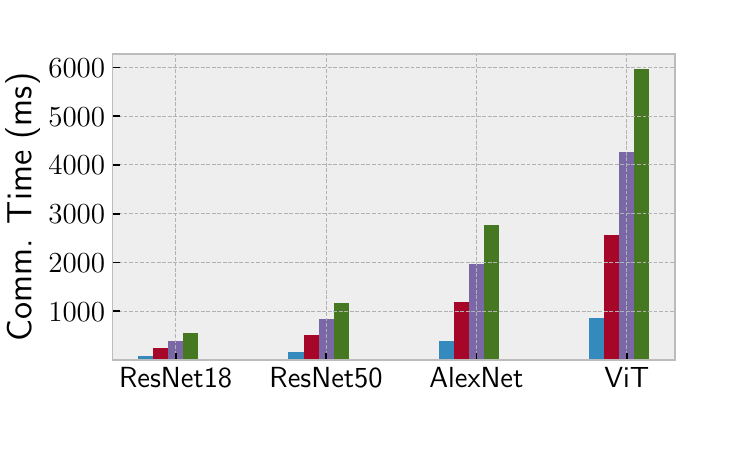}
	\label{fig:allgatherNs}}
	\hfill
	\subfloat[AR-Top\textit{k} communication]{\includegraphics[width=0.23\textwidth]{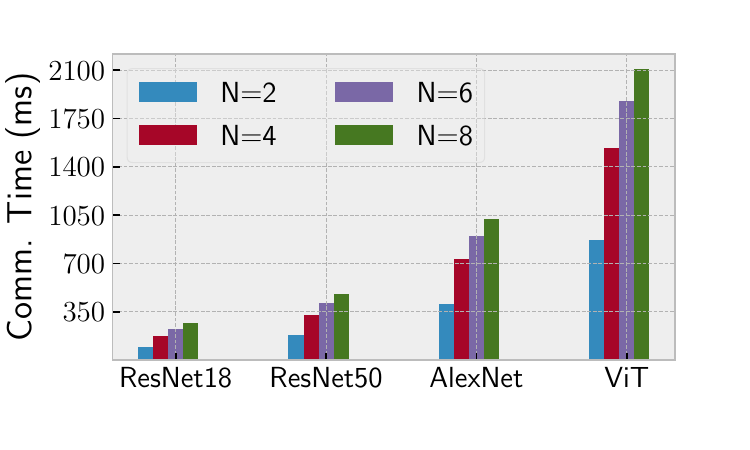}
	\label{fig:artopkNs}}
	\caption{Communication time increases more steeply with $N$ in Allgather vs. AR-Top\textit{k} CR 0.1.
	Network has an average 5ms latency and 1Gbps bandwidth.}
	\label{fig:agartopkwithNs}
\end{figure}

\vspace{0.12cm}
Moreover, we compare the \emph{scale-out} cost between AR-Top\textit{k} and AG as cluster-size increases.
Fig. (\ref{fig:agartopkwithNs}) shows the results for CR 0.1 as $N$ scales from 2 to 8 on a network with an average latency of 5ms and a low 1Gbps bandwidth.
The cost of ART-Ring has a gradual incline, while cost of AG collective increases much more sharply when $N$ is large.
This is because the bandwidth-cost of both broadcast increases as logarithm of $N$, while Ring-AR is near-independent of $N$.
On the other hand, bandwidth cost of AG increases on larger clusters.

\vspace{0.1cm}
As cost of AR-Top\textit{k} can vary based on the AR collective used, we develop a rough heuristic to determine which AR reduction is apt based on current network configuration, model and CR.
Looking at the $\alpha$-$\beta$ costs of ART-Ring and ART-Tree in Eqn. (\ref{eqn:comcostartopk}), the former should be used if its cost is lower than the latter, i.e., $t_{ring}^{(c)} < t_{tree}^{(c)}$.
From solving, ART-Ring should be used over ART-Tree if the $\alpha$-$\beta$ relationship follows Eqn. (\ref{eqn:ringtreecondition}).
Additionally, we earlier saw that at times its better to use AG for communication over AR-Top\textit{k}.
For a DNN with $M$ bytes of gradients compressed to CR $c$, ART-Ring or ART-Tree should be used over AG if the latency-bandwidth ratio satisfies Eqn. (\ref{eqn:artopkringAG}) or (\ref{eqn:artopktreeAG}) respectively, otherwise communication is faster with AG collective.

\begin{subequations}
	\begin{equation}
		(\dfrac{\alpha}{\beta})_{\text{ART-Ring\_over\_ART-Tree}} < \Big(\dfrac{\log (N) - (N-1)/N}{N-1-\log (N)}\Big) Mc
		\label{eqn:ringtreecondition}
	\end{equation}
	\begin{equation}
		(\dfrac{\alpha}{\beta})_{\text{ART-Ring\_over\_AG}} < \Big(1 - \dfrac{1}{N} - \dfrac{\log (N)}{2(N-1)}\Big) Mc
		\label{eqn:artopkringAG}
	\end{equation}
	\begin{equation}
		(\dfrac{\alpha}{\beta})_{\text{ART-Tree\_over\_AG}} < \Big(\dfrac{(N-1)}{\log (N)} - \dfrac{3}{2}\Big) Mc
		\label{eqn:artopktreeAG}
	\end{equation}
	\label{eqn:commcostartopkag}
\end{subequations}

\subsection{\textbf{Compression as a Multi-Objective Optimization Problem}}\label{sec:needflexiblecompress}

The parallel and statistical efficiency of gradient compression with distributed training is pareto-related, i.e., one improves at the detriment of the other.
This is seen from the step-times and accuracy numbers in both Tables (\ref{table:lwmstopkconvergence}) and (\ref{table:artopkconvergence})) where smaller CRs lead to lower step-times (thus improving parallel efficiency) but at the cost of lower final accuracy (degrading the statistical efficiency).
As gradient sensitivity changes throughout training, it is intuitive to change CR as training progresses as well.
For e.g., we could use high CRs in early stages when gradients are volatile and lower CRs later when gradients have saturated and stop changing aggressively.
In order to accelerate training while still achieving DenseSGD-level convergence, we propose an adaptive compression scheme on top of AR-Top\textit{k} that models compression as a \emph{\textbf{multi-objective optimization (MOO)}} problem by identifying three key metrics that affect the pareto-relationship between parallel and statistical efficiency:

\begin{enumerate}
	\item \emph{\textbf{Compression time}:} To first compress and then decompress the gradients, and depends on the compression method itself and may vary with CR.
	E.g., top-\textit{k} sorting based on max-heap for $G$ gradients has complexity $\mathcal{O}(G + k\log G)$.
	In this case, time decreases with CR.
	\item \emph{\textbf{Communication time}:} Between AG, ART-Ring and ART-Tree, the collective with least communication cost depends on $\alpha$, $\beta$, $M$, $N$ and CR (described in Eqn. (\ref{eqn:commcostartopkag})).
	\item \emph{\textbf{Compression Gain}:} Depends on factors like compression technique, CR, stage of training, and indicates statistical efficiency by measuring gradient information lost in compression.
\end{enumerate}




\begin{equation}
	c_{optimal} = \operatorname*{argmin}_c \{\mathcal{F}(\:t_{comp}^{(c)}, t_{sync}^{(c)}, gain^{-1(c)})\}
	\label{eqn:minimizeCR}
\end{equation}

The optimal CR in Eqn. (\ref{eqn:minimizeCR}), $c_{optimal}$ can be determined as a solution to a MOO problem that aims to minimize compression time ($t_{comp}$) and communication time ($t_{sync}$), while maximizing compression $gain$ (or minimizing $1/gain$).
As both network parameters and gain can vary throughout training, so does $c_{optimal}$.
Here, we bound candidate CRs in a min-max exploration space [$c_{low}, c_{high}$] so as to avoid losing important updates at low CRs or incur considerable communication cost at high CRs.
The search for ideal CR in the current state is triggered whenever either the average latency or bandwidth changes beyond a certain threshold.
A background process measures bandwidth using \texttt{iperf} and latency via \texttt{traceroute} between the nodes.
The same process also uses \texttt{tc} to control $\alpha$ and $\beta$ to emulate different network settings in our evaluation as described later.
Based on $\alpha$, $\beta$, $M$, $N$ and CR $c$, the collective with least communication overhead is chosen (by Eqn. (\ref{eqn:commcostartopkag})).
Compression gains for candidate CRs are re-evaluated only in sensitive periods if the inter-interation gain with the current CR changes beyond a specified threshold.
To do this, we first preserve the current model state via checkpoint-restore, select a candidate CR and run it for a few iterations, log the average gain and compression time, and then reload the saved state from checkpoint to evaluate the next candidate CR.
The checkpoint-restore technique ensures model quality is not degraded from evaluating very low CRs.


 \begin{figure}[t]
	 \subfloat[Network Configuration $C1$]{\includegraphics[width=0.23\textwidth]{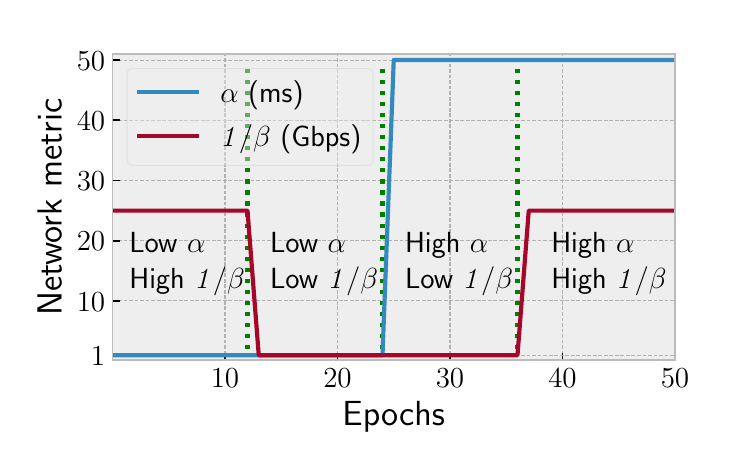}
	\label{fig:networksimall3}}
	\hfill
	\subfloat[Network Configuration $C2$]{\includegraphics[width=0.23\textwidth]{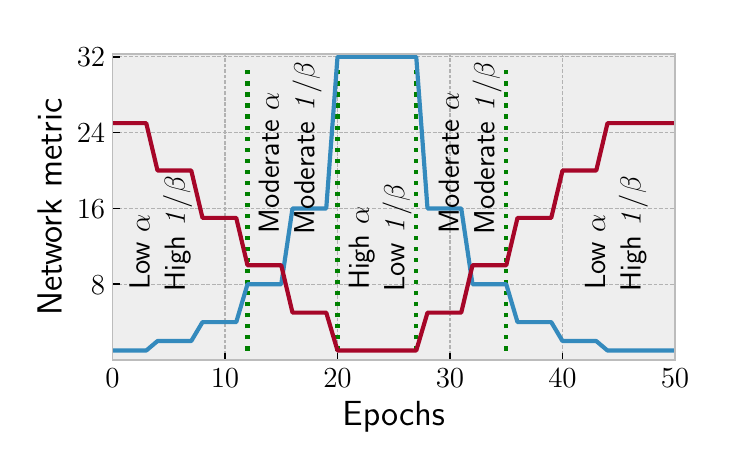}
	\label{fig:networksimresnet50}}
	\caption{Latency and Bandwidth is adjusted via `traffic control' over the epochs to emulate different networking scenarios of low/moderate/high $\alpha$ and $\beta$.}
	\label{fig:networksimmoo}
\end{figure}

\vspace{0.25cm}
\subsubsection{\textit{MOO Configuration Details}}

We assign $c_{low}$ = 0.001, $c_{high}$ = 0.1, and set candidate CRs by scaling $c_{low}$ with a factor of 3.
Thus, selected candidate CRs are [0.1, 0.033, 0.011, 0.004, 0.001], where each one is launched for only 10 iterations in order to measure its $gain$, $t_{comp}$ and $t_{sync}$ to formulate the MOO problem.
The overhead of this exploration phase is low as its trained for only a few iterations and checkpoint-restore is performed in system memory (thus avoiding expensive disk read/writes).
Additionally, this evaluation is triggered only when the inter-iteration gain with current CR, or \emph{gain-threshold} changes by 10\% or more.
Following this, we initiate the search for $c_{optimal}$ \emph{only} if the emulated network latency or bandwidth changes.
This is done since changes in $\alpha$ and $\beta$ influence the communication time of CRs (i.e., $t_{sync}$), while other factors like $M$ and $N$ stay the same.
Based on current network and training configuration, the optimal collective with least communication cost is selected.
If AR-Top\textit{k} is chosen over AG, we save the current model state and select ART-Ring or ART-Tree (whichever is faster according to Eqn. (\ref{eqn:ringtreecondition})) by setting \texttt{NCCL\_ALGO} environment variable to \texttt{RING}\:/\:\texttt{TREE} and redeploying the model from the last saved state.
We use the NSGA-II algorithm (Non-dominated Sorting Genetic Algorithm) \cite{b46} on top of a MOO framework called \texttt{pymoo} \cite{b45} to find $c_{optimal}$ using data gathered from candidate CRs.

\vspace{0.1cm}
Fig. (\ref{fig:networksimmoo}) shows configuration $C1$ and $C2$ where each epoch has a specific latency and bandwidth, set using \texttt{tc}.
Specifically, we emulate 4 scenarios in $C1$: (\textit{low-$\alpha$, high-$1/\beta$}) from epoch 1-12, (\textit{low-$\alpha$, low-$1/\beta$}) from 13-24, (\textit{high-$\alpha$, low-$1/\beta$}) from 25-36 and (\textit{high-$\alpha$, high-$1/\beta$}) thereafter.
We regard 1ms as low and 50ms as high-latency in this configuration, while inter-node $1/\beta$ of 25 and 1Gbps is considered high and low bandwidth respectively.
Configuration $C2$ imitates settings with (\textit{moderate-$\alpha$, moderate-$1/\beta$}) between epochs 12-19 and 28-35, (\textit{high-$\alpha$, low-$1/\beta$}) between 20-27 and (\textit{low-$\alpha$, high-$1/\beta$}) between epochs 0-11 and 36 onwards.
This schedule applies to ResNet18, AlexNet and ViT that train for 50 epochs.
For ResNet50 that runs for 100 epochs, we scale the number of epochs for each ($\alpha$, $1/\beta$) configuration by 2$\times$.
This way, $C1$ applies (\textit{low-$\alpha$, high-$1/\beta$}) state from epoch 1-24 (instead of 1-12), etc.
Similarly, $C2$ for ResNet50 exhibits (\textit{high-$\alpha$, low-$1/\beta$}) between epochs 40-54, and so on.

\begin{figure}[t]
	\centering 
	\subfloat[ResNet18]{\includegraphics[width=0.23\textwidth]{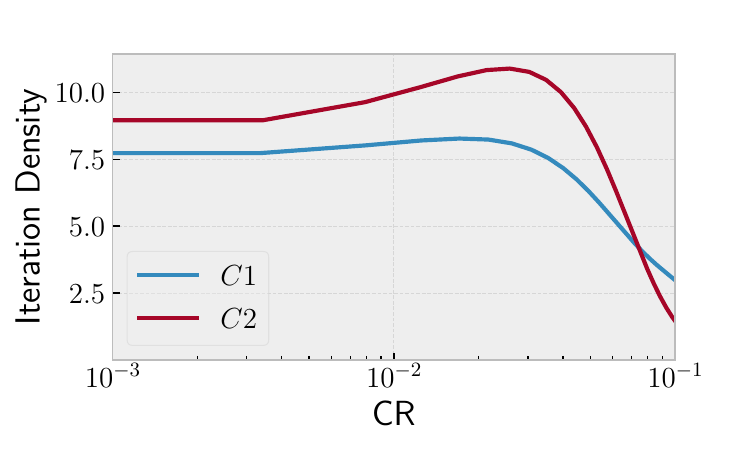}
	\label{resnet18moocr}}
	\subfloat[ResNet50]{\includegraphics[width=0.23\textwidth]{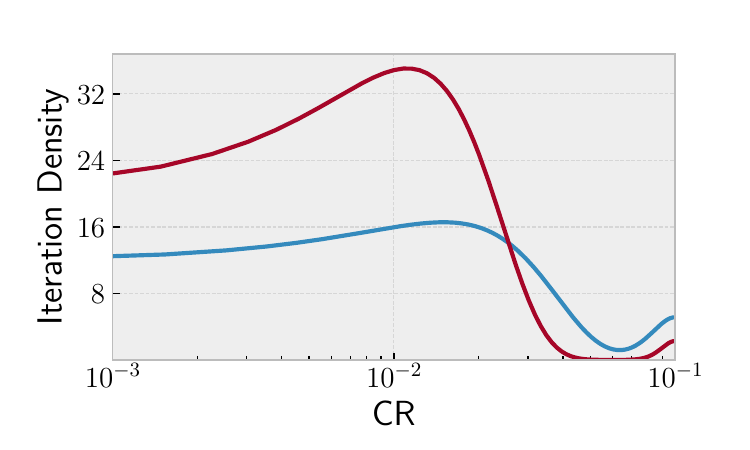}
	\label{resnet50moocr}}
	\hspace{0.01cm}
	\subfloat[AlexNet]{\includegraphics[width=0.23\textwidth]{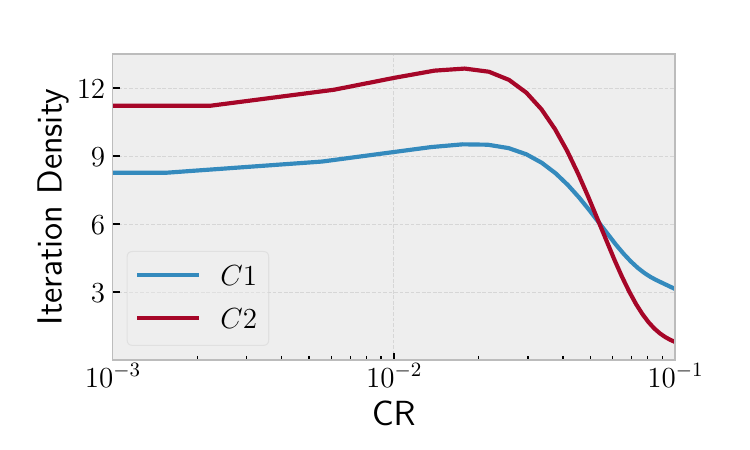}
	\label{alexnetmoocr}}
	\subfloat[ViT]{\includegraphics[width=0.23\textwidth]{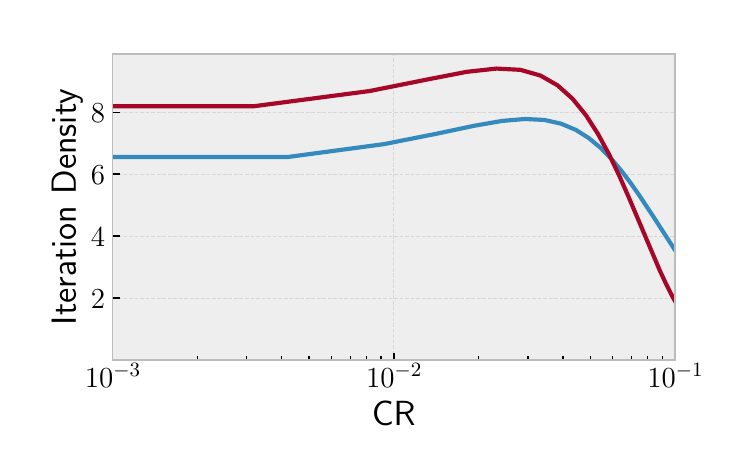}
	\label{vitmoocr}}
	\caption{Iteration density distributions of different CRs used over the course of training for $C1$ and $C2$ configurations when gain-threshold is set to 10\%.}	
	\label{fig:moccompCRs}
\end{figure}

\vspace{0.3cm}
\subsubsection{\textit{Performance of MOO-based Compression:}}

\textit{For $C1$ and $C2$, ResNet18 reached 89.88\% and 90.1\% accuracy, while ResNet50 converged to 98.56\% and 98.62\% respectively.
AlexNet reached 82.4\% in $C1$ and 83.6\% with $C2$, while ViT converged to 80.75\% and 81.2\% with $C1$ and $C2$ respectively.}

\vspace{0.1cm}
By comparing with convergence on \emph{static} CRs in AR-Top\textit{k} (Table (\ref{table:artopkconvergence})) or AG (LW\:/\:MS-Top\textit{k} in Table (\ref{table:lwmstopkconvergence})), we see that dynamically adjusting CR during training while cognizant of a model's $compression$ $gain$ achieves better test accuracy.
In the larger AlexNet and ViT models, MOO-based training achieves slightly better accuracy than even DenseSGD as flexible CRs allow for better exploration of the local minimas on individual workers and thus, improves generalization.
Fig. (\ref{fig:moccompCRs}) plots the kernel density estimates (KDE) of training iterations with various CRs used throughout training for $C1$ and $C2$.
The CRs selected via MOO are the knee-point or pareto-front of compression time, communication time and gain, and lie in the min-max range [$c_{low}$, $c_{high}$].
$C2$ has higher density than $C1$ as search for $c_{optimal}$ is triggered only when network latency or bandwidth changes over the epochs, and $C2$ changes much more often than $C1$ (from Fig. (\ref{fig:networksimresnet50})).
The density peaks lie between CR 0.01 and 0.1 in Fig. (\ref{fig:moccompCRs}), implying that most iterations use a CR in that range for most of the training.


\begin{figure}[t]
	\centering 
	\subfloat[ResNet18]{\includegraphics[width=0.23\textwidth]{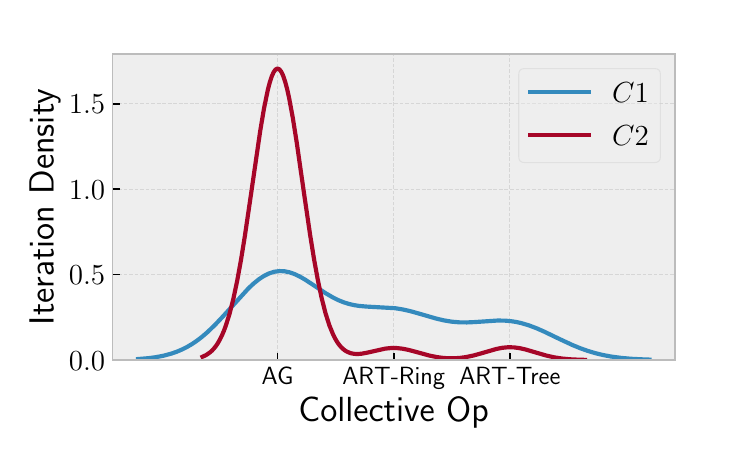}
	\label{resnet18moocollect}}
	\subfloat[ResNet50]{\includegraphics[width=0.23\textwidth]{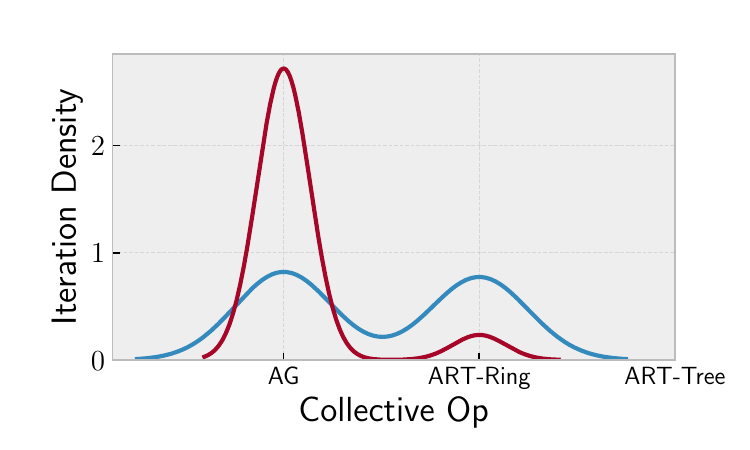}
	\label{resnet50moocollect}}
	\hspace{0.01cm}
	\subfloat[AlexNet]{\includegraphics[width=0.23\textwidth]{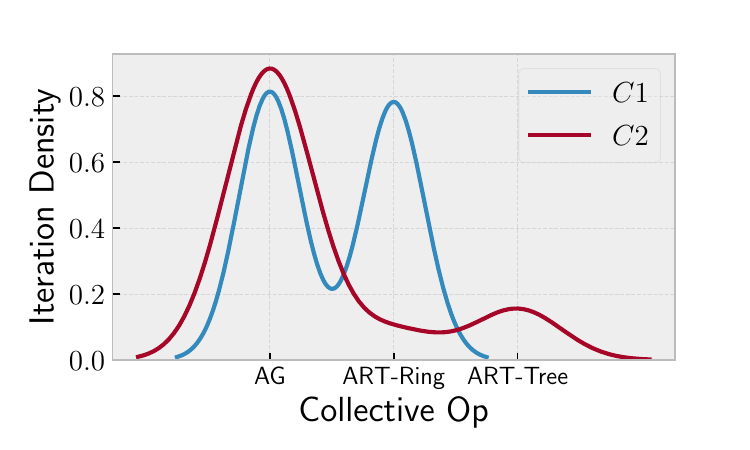}
	\label{alexnetmoocollect}}
	\subfloat[ViT]{\includegraphics[width=0.23\textwidth]{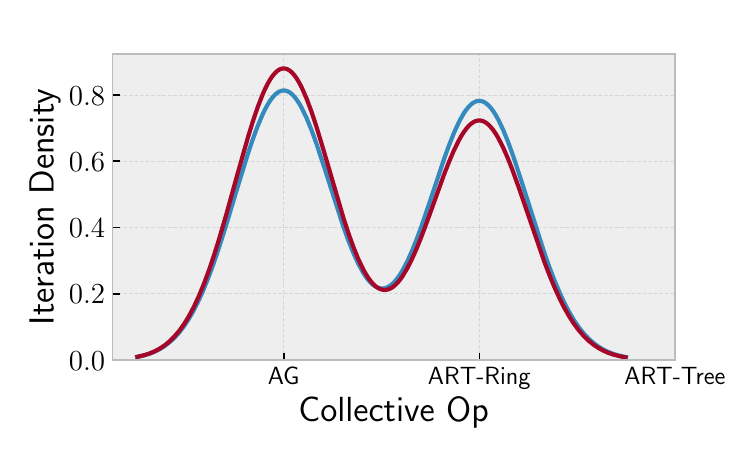}
	\label{vitmoocollect}}
	\caption{Density distributions of AG, ART-Ring and ART-Tree collectives used throughout training for configurations $C1$ and $C2$ with gain-threshold 10\%.}
	\label{fig:moccompcollectives}
\end{figure}

\vspace{0.1cm}
Based on $\alpha$-$\beta$ cost model, collectives have varying overheads depending on network configuration, CR, model and cluster-size.
While optimizing for $c_{optimal}$, we dynamically choose the collective with least communication cost as well.
Fig. (\ref{fig:moccompcollectives}) shows the iteration densities of different collectives used over the course of training as the network configuration changes.
For smaller models like ResNet18 and ResNet50, configuration $C2$ uses AG over AR-Top\textit{k} for most iterations.
A major chunk of $C2$ comprises of (\textit{low}-$\alpha$, \textit{high}-$1/\beta$) and \textit{moderate}-($\alpha$, $1/\beta$) phase.
In Table (\ref{table:simcommartopkAG}), we see that both ResNet18 and ResNet50 use AG at CRs 0.01 and 0.001 and a 1ms, 10Gbps network; a low-latency, moderate-to-high bandwidth network setting.
For ResNet50 and ViT, ART-Tree always had a higher synchronization overhead than ART-Ring.
Thus, density of ART-Tree is 0 in Fig. (\ref{resnet50moocollect}) and (\ref{vitmoocollect}).
Larger models like AlexNet and ViT communicate with both AG and ART-Ring, as similarly seen from the communication time for different $\alpha$-$\beta$ settings in Table (\ref{table:simcommartopkAG}).

\section{Discussion}

Unlike conventional distributed applications, not all work performed in DNN training holds equal contribution towards final model quality due to stochastic nature of SGD and critical periods in the learning phase.
This opens up new avenues of adaptive communication for training speedups \cite{b1, b14, b2400}.
Gradient compression trade-offs speedup with convergence quality based on the chosen compression ratio.
We model the same as a multi-objective optimization problem to minimize compression-compression cost (parallel efficiency), while maximizing compression gain (statistical efficiency) in distributed learning.
Additionally, the optimal collective to use varies with a DNN and latency-bandwidth of a network.

\vspace{0.1cm}
The latter may fluctuate in data-center/edge due to network congestion, contention, scheduling, etc.
Our work predicts the ideal compression configuration to use for fast learning over such unpredictable networks.
Further, this work can improve convergence on unbalanced and skewed data in federated learning with its low-volume, high-frequency communication strategy (thus, routinely sharing important model features) compared to algorithms like FedAvg \cite{b2500} that use a high-volume, low-frequency approach instead.
An ML practitioner may not necessarily be aware of the network topology when deploying models over a cluster of nodes, or the network configuration may be different as the training infrastructure changes (for e.g., training in a private cluster vs. the cloud).
Our system keeps track of the current network latency and available bandwidth to determine the ideal collective op with least communication cost based on the $\alpha$-$\beta$ model.
Apart from top-\textit{k}, our approach is compatible with other compressors (like DGC \cite{b47}, SIDCo \cite{b48}, etc.), and can be replaced easily.
	\vspace{0.15cm}
\section{conclusion and future work}

The optimal communication collective depends on factors like network topology, latency, bandwidth, model-size, cluster-size and degree of compression (if any), among others.
In this work, we propose a new compression-communication mechanism called AR-Top\textit{k} to accelerate training when Allreduce is faster than Allgather.
Further, we model compression as an optimization problem to balance the parallel and statistical efficiency, and see how CR changes with network state and stage of training.
We dynamically choose the CR to attain better convergence quality than with static compression, and achieve maximal speedup with the appropriate communication collective.
Currently, we propose two worker selection mechanisms in AR-Top\textit{k}, one staleness-based and other variance-based.
In our evaluation we see that STAR-Top\textit{k} usually performs better than VAR-Top\textit{k} on small-to-medium cluster-size.
However, the former may degrade model convergence on large cluster-sizes due to stale updates from choosing workers in a round-robin fashion.
In the future, we plan to combine the two approaches where AR-Top\textit{k} automatically switches between the two based on the DNN test performance with each approach.

We also plan to deploy large language models (LLMs) with AR-Top\textit{k} and evaluate trade-offs between convergence and overall speedup over massive attention-based models.
As part of integration with other distributed training paradigms, we expect our adaptive compression technique to work even with model-parallelism, where gradient updates between subsequent layers can be compressed to different CRs over the course of training.



\begin{thebibliography}{00}
		\bibitem{b0} Li, Mu et al. “Scaling Distributed Machine Learning with the Parameter Server.” BigDataScience '14 (2014).
		\bibitem{b1} Tyagi, Sahil et al. “GraVAC: Adaptive Compression for Communication-Efficient Distributed DL Training.” IEEE CLOUD (2023).
		\bibitem{b2} Agarwal, Saurabh et al. “On the Utility of Gradient Compression in Distributed Training Systems.” MLSys (2021).
		\bibitem{b3} McCandlish, Sam et al. “An Empirical Model of Large-Batch Training.” ArXiv abs/1812.06162 (2018).
		\bibitem{b4} Tyagi, Sahil et al. “Scavenger: A Cloud Service for Optimizing Cost and Performance of ML Training.” IEEE/ACM CCGrid (2023).
		\bibitem{b5} Qiao, Aurick et al. “Pollux: Co-adaptive Cluster Scheduling for Goodput-Optimized Deep Learning.” OSDI (2020).
		\bibitem{b6} NVLink: https://www.nvidia.com/en-us/data-center/nvlink/.
		\bibitem{b7} NVIDIA DGX: https://www.nvidia.com/en-us/data-center/dgx-systems/.
		\bibitem{b9} AWS https://docs.aws.amazon.com/AWSEC2/latest/UserGuide/ec2-instance-network-bandwidth.html.
		\bibitem{b11} Sun, Peng et al. “Optimizing Network Performance for Distributed DNN Training on GPU Clusters: ImageNet/AlexNet Training in 1.5 Minutes.” ArXiv abs/1902.06855 (2019).
		\bibitem{b14} Tyagi, Sahil and Swany, Martin. “Accelerating Distributed ML Training via Selective Synchronization.” IEEE CLUSTER (2023).
		\bibitem{b16} Li, Shen et al. “PyTorch distributed.” VLDB Endowment 13 (2020).
		\bibitem{b18} Horovod Tensor Fusion: https://horovod.readthedocs.io/en/stable/tensor-fusion\_include.html.
		\bibitem{b19} Mai, Luo et al. “KungFu: Making Training in Distributed Machine Learning Adaptive.” USENIX OSDI (2020).
		\bibitem{b20} Johnson, Tyler B. et al. “AdaScale SGD: A User-Friendly Algorithm for Distributed Training.” ArXiv abs/2007.05105 (2020).
		\bibitem{b21} Frankle, Jonathan et al. “The Early Phase of Neural Network Training.” ArXiv abs/2002.10365 (2020).
		\bibitem{b22} Achille, Alessandro et al. “Critical Learning Periods in Deep Neural Networks.” ArXiv abs/1711.08856 (2017).
		\bibitem{b23} Agarwal, Saurabh et al. “Accordion: Adaptive Gradient Communication via Critical Learning Regime Identification.” abs/2010.16248 (2020).
		\bibitem{b24} Yao, Zhewei et al. “Large batch training of neural networks with adversarial training and second-order information.” abs/1810.01021 (2018).
		\bibitem{b25} Alistarh, Dan et al. “The Convergence of Sparsified Gradient Methods.” NIPS (2018).
		\bibitem{b26} Shi, Shaohuai et al. “Towards Scalable Distributed Training of Deep Learning on Public Cloud Clusters.” ArXiv abs/2010.10458 (2020).
		\bibitem{b2400} Tyagi, S., and Swany, M. (2022). ScaDLES: Scalable Deep Learning over Streaming data at the Edge. 2022 IEEE Big Data, 2113-2122.
		\bibitem{b2500} McMahan, H. B. et al. “Communication-Efficient Learning of Deep Networks from Decentralized Data.” AISTATS (2016).
		\bibitem{b27} Bernstein, Jeremy et al. “signSGD: compressed optimisation for non-convex problems.” ArXiv abs/1802.04434 (2018).
		\bibitem{b28} Wen, Wei et al. “TernGrad: Ternary Gradients to Reduce Communication in Distributed Deep Learning.” NIPS (2017).
		\bibitem{b29} Seide, Frank et al. “1-bit stochastic gradient descent and its application to data-parallel distributed training of speech DNNs.” Interspeech (2014).
		\bibitem{b30} Vogels, Thijs et al. “PowerSGD: Practical Low-Rank Gradient Compression for Distributed Optimization.” NIPS (2019).
		\bibitem{b31} Sarvotham, Shriram et al. “Connection-level analysis and modeling of network traffic.” International Memory Workshop (2001).
		\bibitem{b32} He, Kaiming et al. “Deep Residual Learning for Image Recognition.” 2016 IEEE Conference on Computer Vision and Pattern Recognition (CVPR) (2015): 770-778.
		\bibitem{b33} Krizhevsky, Alex et al. “ImageNet classification with deep convolutional neural networks.” Communications of the ACM 60 (2012): 84 - 90.
		\bibitem{b34} Dosovitskiy, Alexey et al. “An Image is Worth 16x16 Words: Transformers for Image Recognition at Scale.” ArXiv abs/2010.11929 (2020).
		\bibitem{b35} Krizhevsky, Alex. “Learning Multiple Layers of Features from Tiny Images.” (2009).
		\bibitem{b37} GCP: https://cloud.google.com/compute/docs/networking/configure-vm-with-high-bandwidth-configuration.
		\bibitem{b38} Egress BW: https://cloud.google.com/compute/docs/network-bandwidth.
		\bibitem{b39} Linux traffic control: https://man7.org/linux/man-pages/man8/tc.8.html.
		\bibitem{b40} Fei-Fei, Li et al. “Learning Generative Visual Models from Few Training Examples: An Incremental Bayesian Approach Tested on 101 Object Categories.” CVPR Workshop (2004).
		\bibitem{b41} G. Griffin, et al.“Caltech 256”. doi: 10.22002/D1.20087.
		\bibitem{b42} Thakur, Rajeev et al. “Optimization of Collective Communication Operations in MPICH.” IJHPCA (2005).
		\bibitem{b43} Bossard, Lukas et al. “Food-101 Mining Discriminative Components with Random Forests.” ECCV (2014).
		\bibitem{b44} iPerf test tool for TCP, UDP and SCTP: https://iperf.fr/.
		\bibitem{b45} Blank, Julian and Kalyanmoy Deb. “Pymoo: Multi-Objective Optimization in Python.” IEEE Access 8 (2020): 89497-89509.
		\bibitem{b46} Deb, Kalyanmoy, Samir Agrawal, Amrit Pratap and T. Meyarivan. “A fast and elitist multiobjective genetic algorithm: NSGA-II.” IEEE Trans. Evol. Comput. 6 (2002): 182-197.
		\bibitem{b47} Lin, Yujun et al. “Deep Gradient Compression: Reducing the Communication Bandwidth for Distributed Training.” ICLR (2018).
		\bibitem{b48} Abdelmoniem, Ahmed Mohamed et al. “An Efficient Statistical-based Gradient Compression Technique for Distributed Training Systems.” MLSys (2021).
	\end{thebibliography}
\end{document}